\pdfoutput=1
\documentclass{ws-ijmpa}
\usepackage[super,compress]{cite}
\usepackage{graphicx}
\begin{document}
\markboth{P.Kokkas}
{Measurements of jet-related observables at the LHC}

\providecommand{\GeV} {\textrm{GeV}\,}
\providecommand{\TeV} {\textrm{TeV}\,}
\providecommand{\pt} {\ensuremath{p_{\textrm{T}}}\,}
\providecommand{\antikt} {\ensuremath{\textrm{anti-}k_{\textrm{T}}}\,}

\providecommand{\ymax}  {\ensuremath{y_{\textrm{max}}}\,}
\providecommand{\aymax} {\ensuremath{\vert y\vert_{\textrm{max}}}\,}
\providecommand{\mjjj}  {\ensuremath{m_3}\,}

\providecommand{\PYTHIA} {{\textsc{pythia}}\,}
\providecommand{\POWHEG} {{\textsc{powheg}}\,}
\providecommand{\HERWIGPP} {{\textsc{herwig++}}\,}
\providecommand{\HERWIG} {{\textsc{herwig}}\,}
\providecommand{\PYTHIA} {{\textsc{pythia}}\,}
\providecommand{\PYTHIAS} {{\textsc{pythia6}}\,}
\providecommand{\PYTHIAE} {{\textsc{pythia8}}\,}
\providecommand{\MADGRAPH} {{\textsc{MadGraph}}\,}
\providecommand{\ALPGEN} {{\textsc{Alpgen}}\,}
\providecommand{\SHERPA} {{\textsc{Sherpa}}\,}
\providecommand{\CASCADE} {{\textsc{Cascade}}\,}
\providecommand{\HEJ} {{\textsc{Hej}}\,}
\providecommand{\ARIADNE} {{\textsc{Ariadne}}\,}

\catchline{}{}{}{}{}
%
%%%%%%%%%%%%%%%%%%%%%%%%%%%%%%%%%%%%%%%%%%%%%%%%%%%%%%%%%%%%%%%%%%%%

\title{Measurements of jet-related observables at the LHC}

\author{P. KOKKAS}

\address{Physics Department, University of Ioannina,\\
45110 Ioannina, Greece\\
pkokkas@uoi.gr}

\maketitle

\begin{history}
\received{Day Month Year}
\revised{Day Month Year}
\end{history}

\begin{abstract}
During the first years of the LHC operation a large amount of jet data was recorded
by the ATLAS and CMS experiments. In this review several measurements of jet-related 
observables are presented, such as multi-jet rates and cross sections, ratios of jet cross 
sections, jet shapes and event shape observables. All results presented here are
based on jet data collected at a center-of-mass energy of 7 TeV.
Data are compared to various Monte Carlo generators, as well as to theoretical 
next-to-leading-order calculations allowing a test of perturbative Quantum Chromodynamics
in a previously unexplored energy region.

\end{abstract}

\ccode{PACS numbers:}

%\tableofcontents

\section{Introduction}	

After three years of very successful operation of the Large Hadron Collider (LHC), a large 
amount of jet data was recorded both by the ATLAS\cite{ATLAS_detector}  and CMS\cite{CMS_detector} experiments.
Up to now these collaborations published results on jet cross sections, ratios of jet cross sections,  
multi-jet rates and other jet related observables using the data collected during the first two years
at a center-of-mass energy of 7 TeV. 
The goals of these studies are to test perturbative Quantum Chromodynamics (pQCD) in a previously unexplored energy region,
check the Standard Model (SM) predictions at high energy scales, measure and understand the 
main background to many new physics searches, determine the strong coupling constant $\alpha_{s}$ 
and its running, and provide constraints on parton distribution functions (PDFs).

At leading order (LO) in pQCD, jet production in proton-proton (pp)
collisions occurs when two partons interact via the strong force to produce two 
final-state partons. 
Each of the final state particles may subsequently lose energy by emitting other quarks and 
gluons in a process referred to as a parton shower (PS). Finally, the products of the parton 
shower undergo hadronisation and form hadron jets. Events with three or more jets in the final 
state originate from hard-gluon radiation and other higher-order QCD processes.

Classical measurements are those of the inclusive jet \pt and dijet mass cross section, 
which are presented in Refs. \refcite{CMS_incl} to \refcite{ATLAS_incl7TeV} and also discussed in a 
separate review in this volume\cite{Francavilla} .
Here, a review of recent measurements on other jet-related observables at the LHC is 
presented, based on the data collected during 2010 and 2011 at a center-of-mass energy of 7 TeV. 
The review is organized as follows: Section 2 presents results on multi-jet rates and cross sections. 
In section 3 measurements of ratios of jet cross sections are presented.
Section 4 presents results on jet shapes observables. In section 5 measurements of event shapes are presented, 
followed by the conclusions in Section 6.

\section{Multi-jet rates and cross sections}

\begin{figure}[hbtp]  
 \begin{center}
 \includegraphics[width=0.40\textwidth]{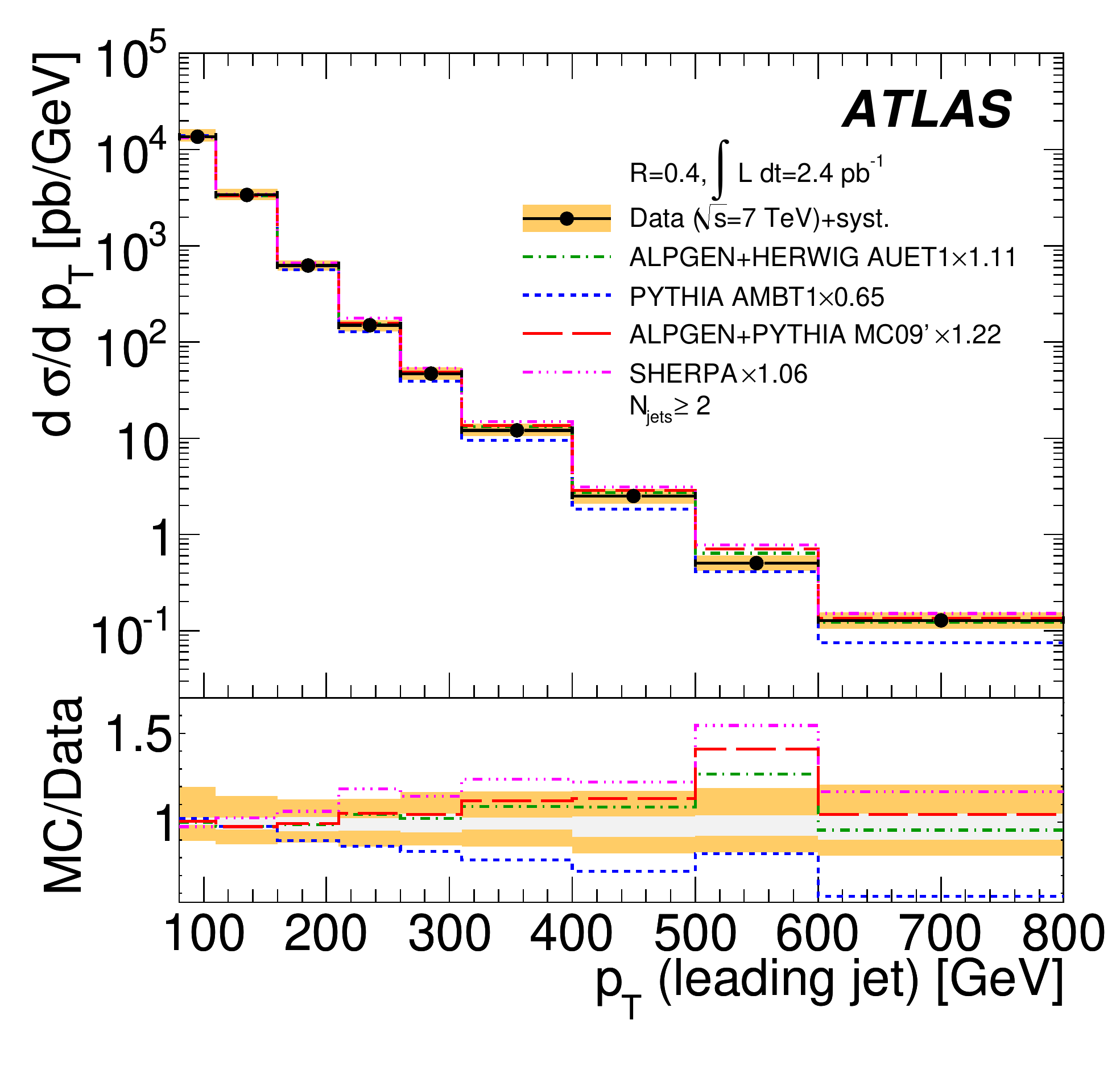}
 \includegraphics[width=0.40\textwidth]{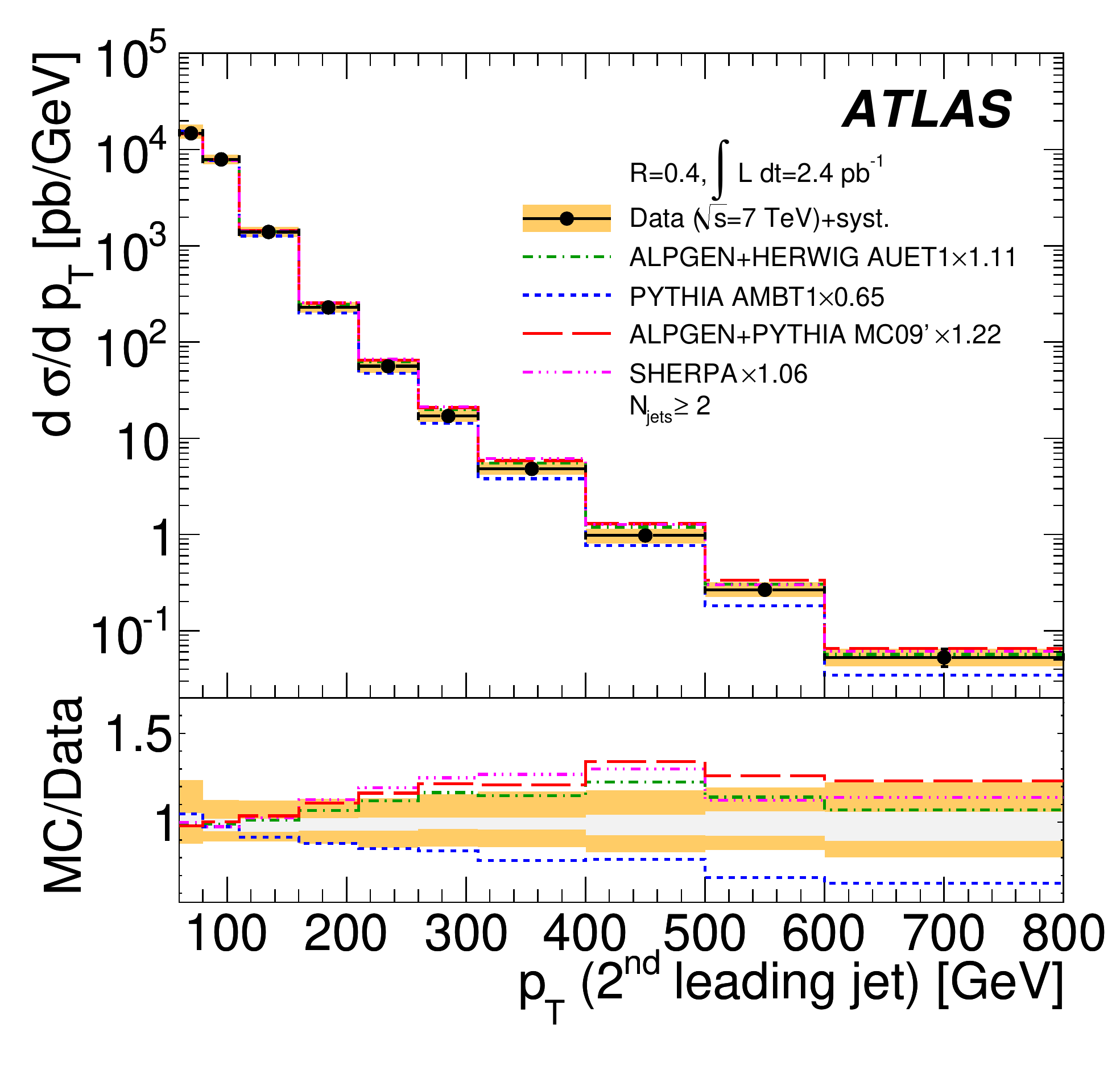}
 \includegraphics[width=0.40\textwidth]{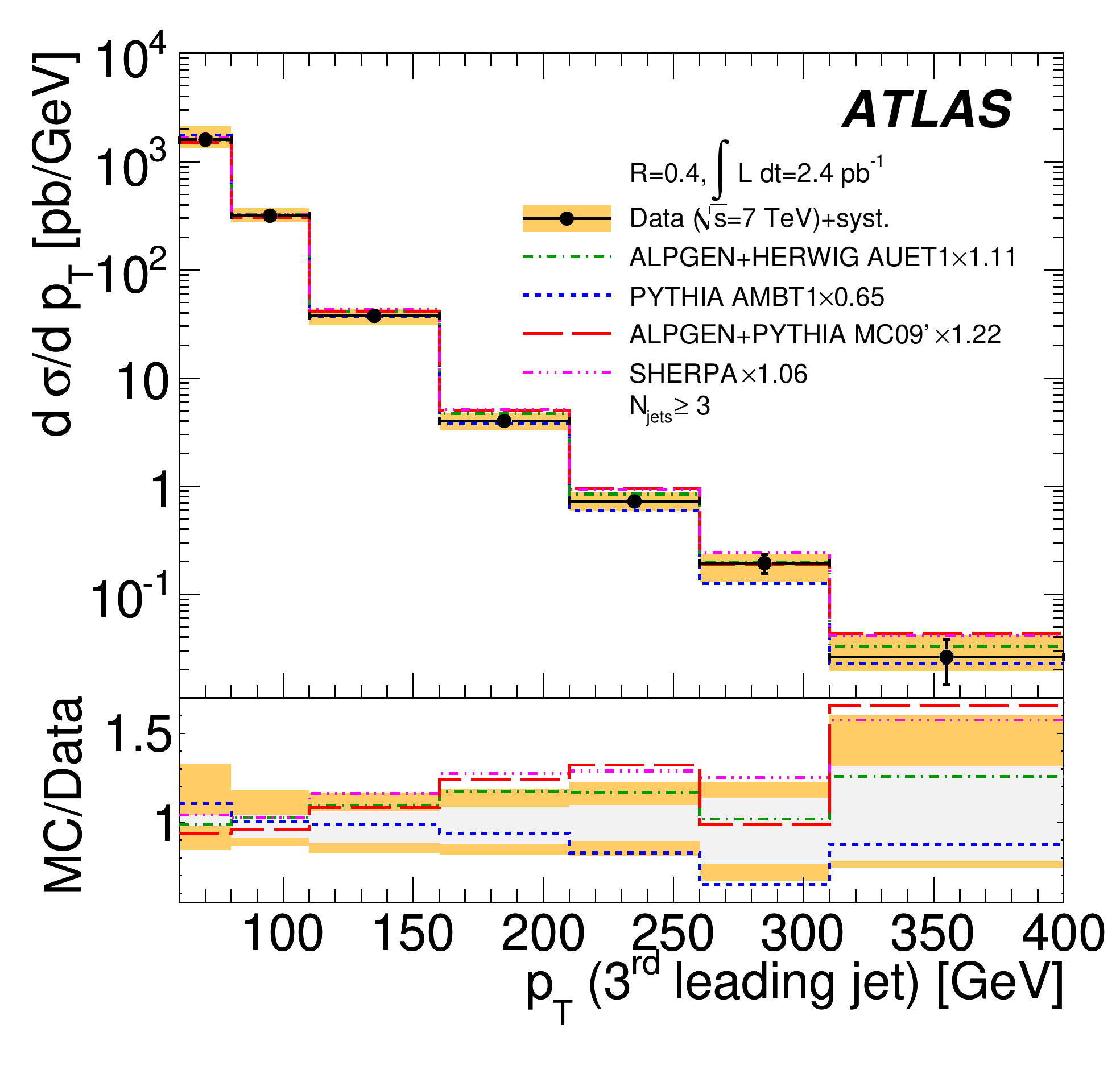}
 \includegraphics[width=0.40\textwidth]{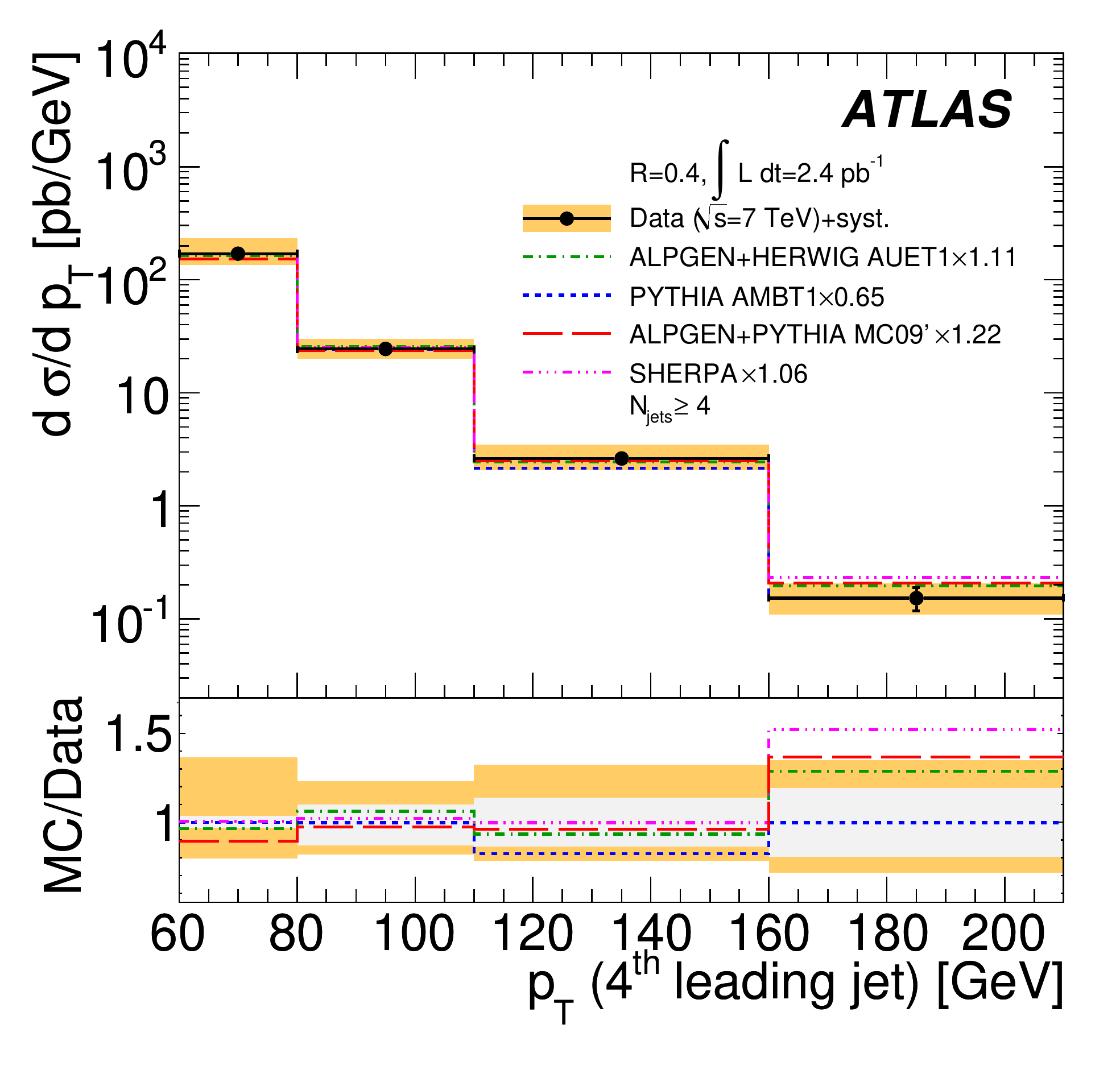}

 \caption{The differential cross section of the four leading jets as a function of jet \pt, 
          measured by the ATLAS collaboration.
          The results are compared to different LO MC simulations.}
 \label{fig:ATLASmultijets}
 \end{center}
\end{figure}

The goal of the studies with the very first LHC jet data was to test the performance of the different 
LO Monte Carlo (MC) simulations, so that they can be used to estimate multi-jet backgrounds for 
new particle searches.
For example the ATLAS collaboration, in Ref.~\refcite{ATLAS_multijets}, presents studies on inclusive 
multi-jet production, using the very first jet data collected during 2010 which correspond to an 
integrated luminosity of $2.4\, \mathrm{pb^{-1}}$.
Figure~\ref{fig:ATLASmultijets} shows the \pt-dependent differential cross sections for the first 
four leading jets in multi-jet events. The results are compared to different LO MC 
simulations based on the \PYTHIA\cite{pythia6} , \ALPGEN\cite{alpgen} and \SHERPA\cite{sherpa} generators using various 
optimised sets of parameters (so-called {\it tunes}). 
All MC simulations agree reasonably well with the data.

\begin{figure}[hbtp]  
 \begin{center}
 \includegraphics[width=0.65\textwidth]{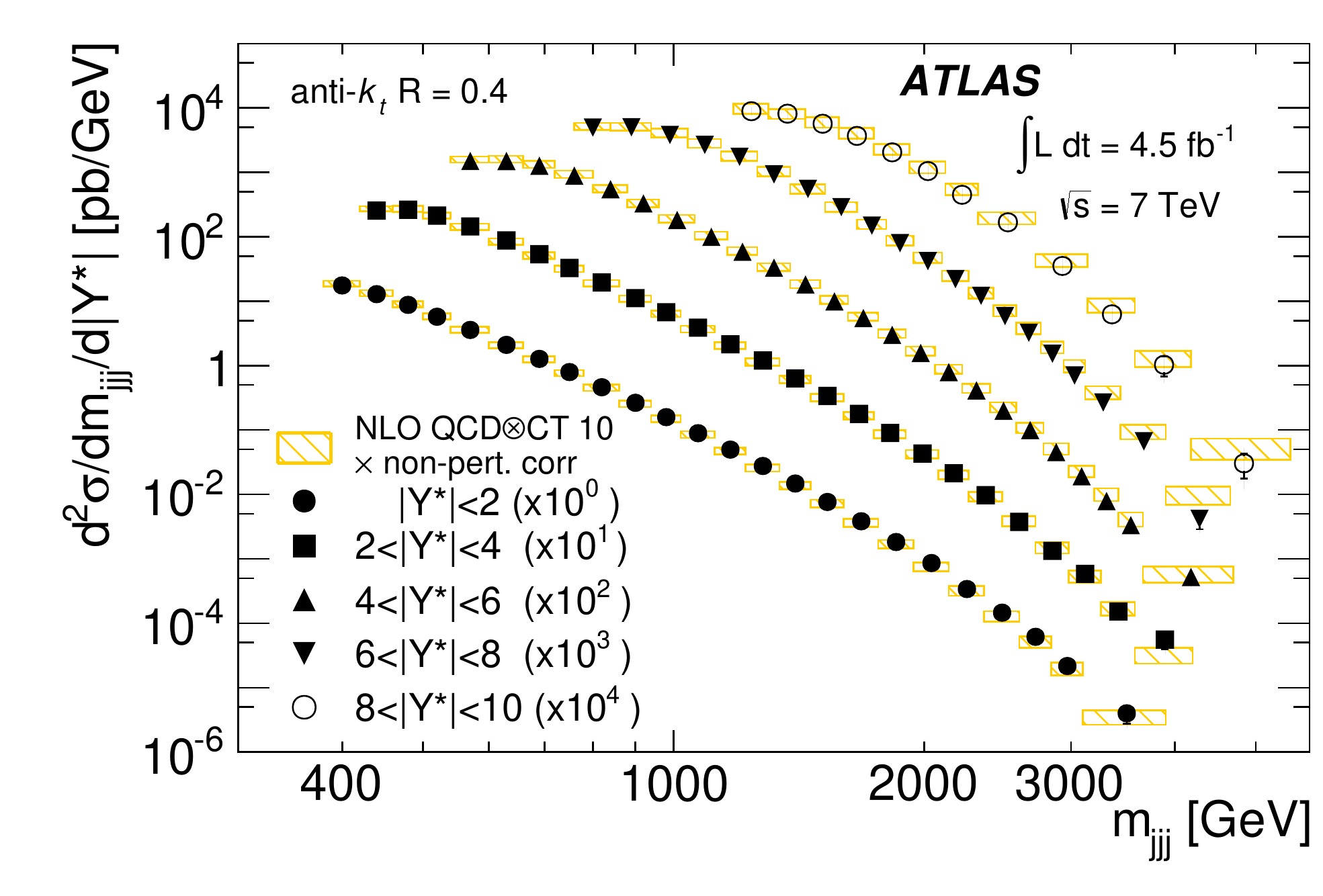}
 \includegraphics[width=0.85\textwidth]{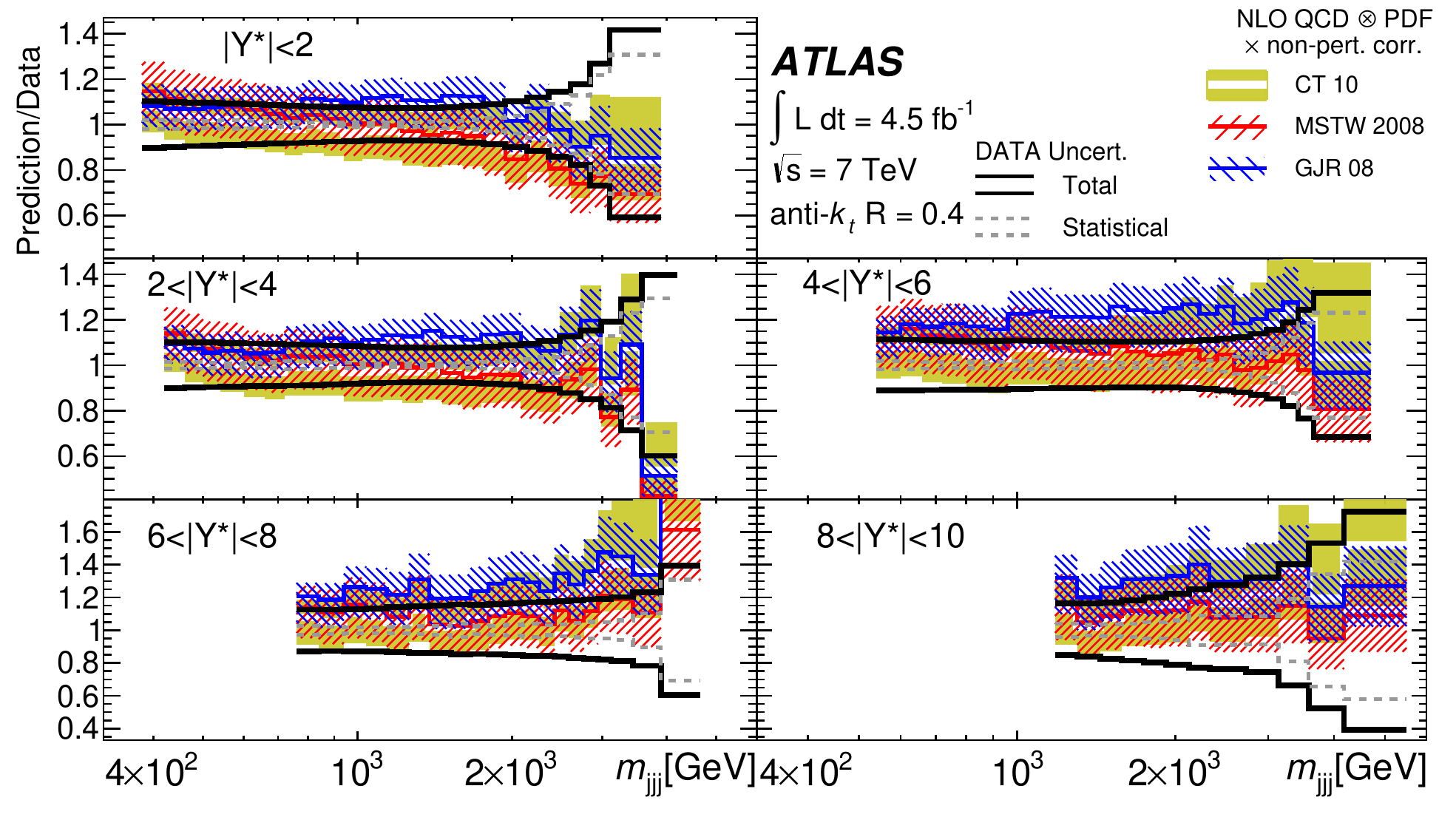}
 \caption{Top: the three-jet double-differential cross section, measured by the ATLAS collaboration, 
          compared to the NLO prediction times NP corrections using the CT10-NLO PDF set.
	  Bottom: the ratio of data to NLO theoretical predictions using various PDF sets.}
 \label{fig:ATLAS_3jetcrossSections}
 \end{center}
\end{figure}

In Ref.~\refcite{ATLAS_3jetCrossSections} the ATLAS collaboration presents the measurement of double-differential 
three-jet production cross sections as a function of the three-jet mass \mjjj and the sum of absolute
rapidity separation between the three leading jets 
$\lvert Y^{*} \rvert=\lvert y_1 - y_2 \rvert + \lvert y_2 - y_3 \rvert + \lvert y_1 - y_3 \rvert$. 
The measurement is done for two different values of the jet radius parameter, $R=0.4$ and $R=0.6$,
using a data sample corresponding to an integrated luminosity of $4.51\, \mathrm{fb^{-1}}$ collected with the ATLAS detector during 2011. 
The goal of this study is to test the description of multi-jet events in next-to-leading-order (NLO) QCD, 
and provide constraints on the proton PDFs beyond those from inclusive and dijet cross sections.
Figure~\ref{fig:ATLAS_3jetcrossSections}, on the top, shows the comparison of the three-jet double-differential 
cross section as a function of \mjjj, binned in $\lvert Y^{*} \rvert$, to NLO predictions times non-perturbative 
(NP) corrections using the CT10\cite{ct10a,ct10b}-NLO PDF set. The same figure on the bottom shows the ratio 
of data to NLO predictions using different PDF sets, like CT10, MSTW2008\cite{mstw2008a,mstw2008b} , and
GJR08\cite{gjr08} . 
It is found that for a jet radius parameter $R=0.4$ most PDF sets are able to describe the data well.
Theory predictions for $R=0.6$ underestimate the data across the full $\mjjj-\lvert Y^{*} \rvert$ plane.
The discrepancy is covered by the experimental and theoretical uncertainty bands and it has only a minor 
dependence on the PDF set used.

\begin{figure}[hbtp]  
 \begin{center}
 \includegraphics[width=0.57\textwidth]{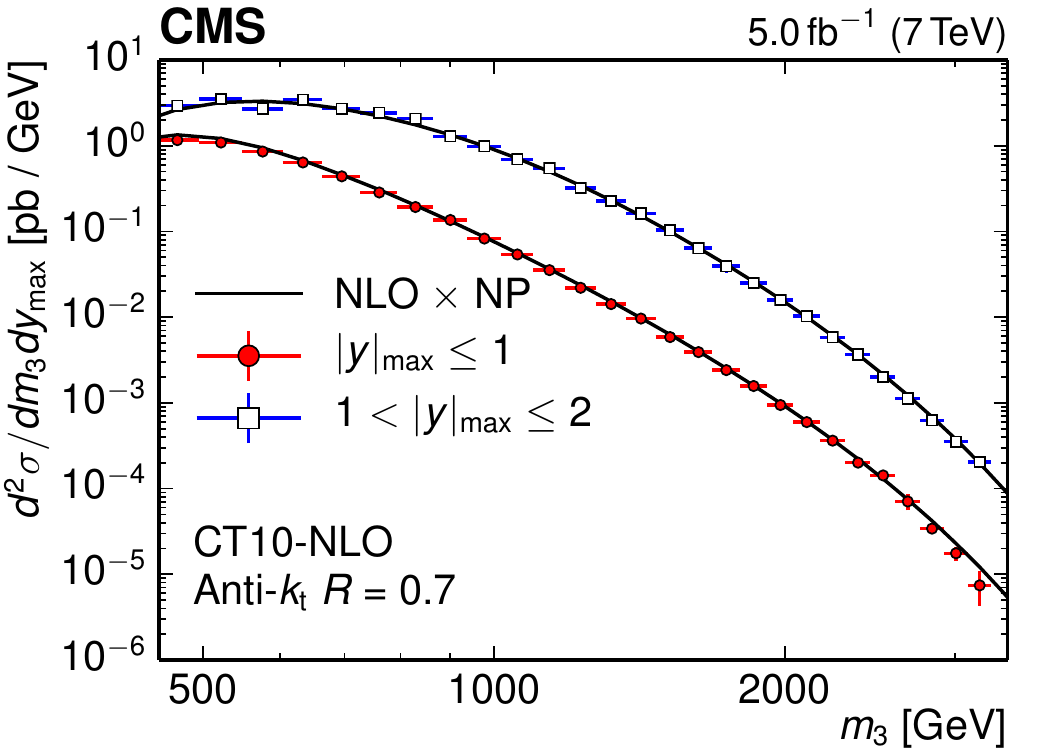}
 \includegraphics[width=0.40\textwidth]{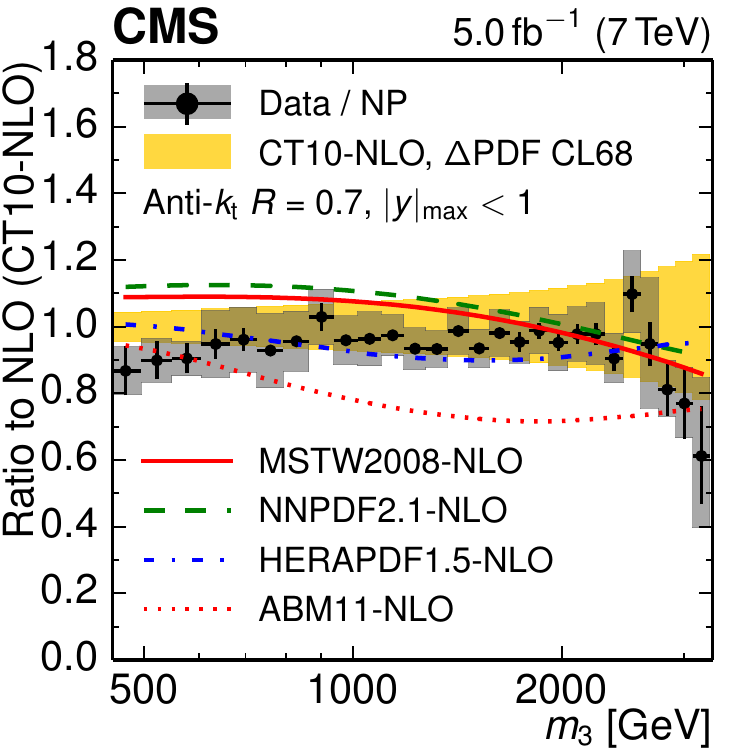}
 \caption{Left: the comparison of the 3-jet mass cross section with the NLO 
	  prediction times NP corrections, for the two considered regions in $\aymax$, 
	  using the CT10-NLO PDF set.
	  Right: ratio of the measured 3-jet mass distribution to theoretical predictions using 
	  various PDF sets with NLO PDF evolution.}
 \label{fig:3jetmass}
 \end{center}
\end{figure}

The differential 3-jet production cross section as a function of the invariant mass \mjjj and the 
maximum rapidity \aymax of the 3-jet system has been published by the CMS collaboration in Ref.~\refcite{CMS_3jetmass}.
The measurement is done in two rapidity bins with $\aymax < 1$ and $1 \leq \aymax < 2$ using an integrated 
luminosity of $5\, \mathrm{fb^{-1}}$ collected with the CMS detector during 2011.
Figure~\ref{fig:3jetmass}, on the left, shows the comparison of the 3-jet mass distribution to the NLO 
prediction employing the CT10 PDF set times non-perturbative corrections. It is observed that pQCD is able to describe 
the cross section as a function of the 3-jet mass over five orders of magnitude and for 3-jet masses up to 3 TeV. 
The same figure on the right presents the ratios of the measured cross sections, for the lowest rapidity 
bin, to theoretical predictions including NP effects, using various PDF sets, such as CT10, NNPDF2.1\cite{nnpdf1,nnpdf2} , 
MSTW2008, HERAPDF1.5\cite{herapdf} and ABM11\cite{abm11} .
Within uncertainties most PDF sets are able to describe the data.
This measurement is used by the CMS collaboration for the determination of the strong coupling constant, $\alpha_{s}$,
see Ref.~\refcite{Juan}.

\begin{figure}[hbtp]  
 \begin{center}
 \includegraphics[width=0.34\textwidth]{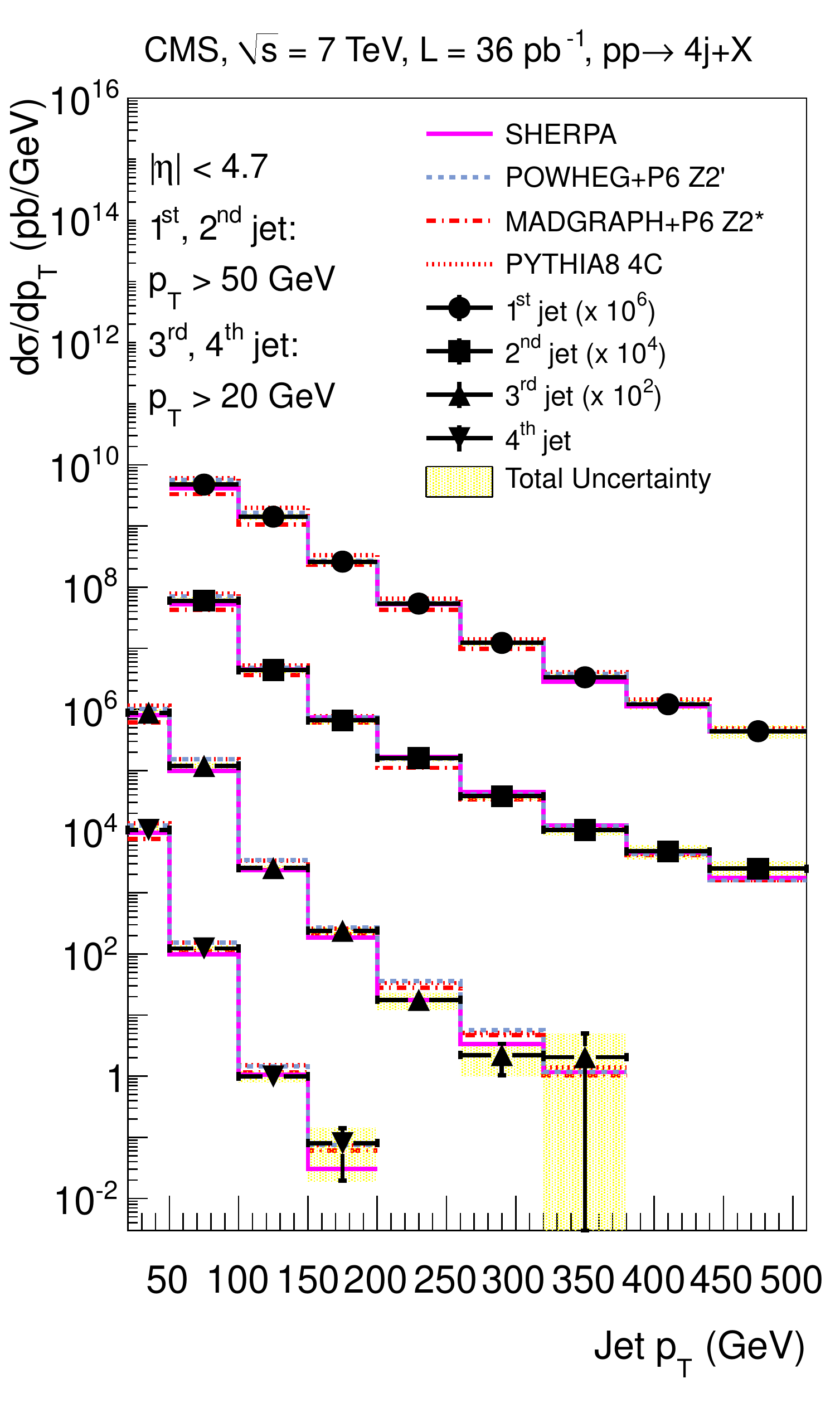}
 \includegraphics[width=0.37\textwidth]{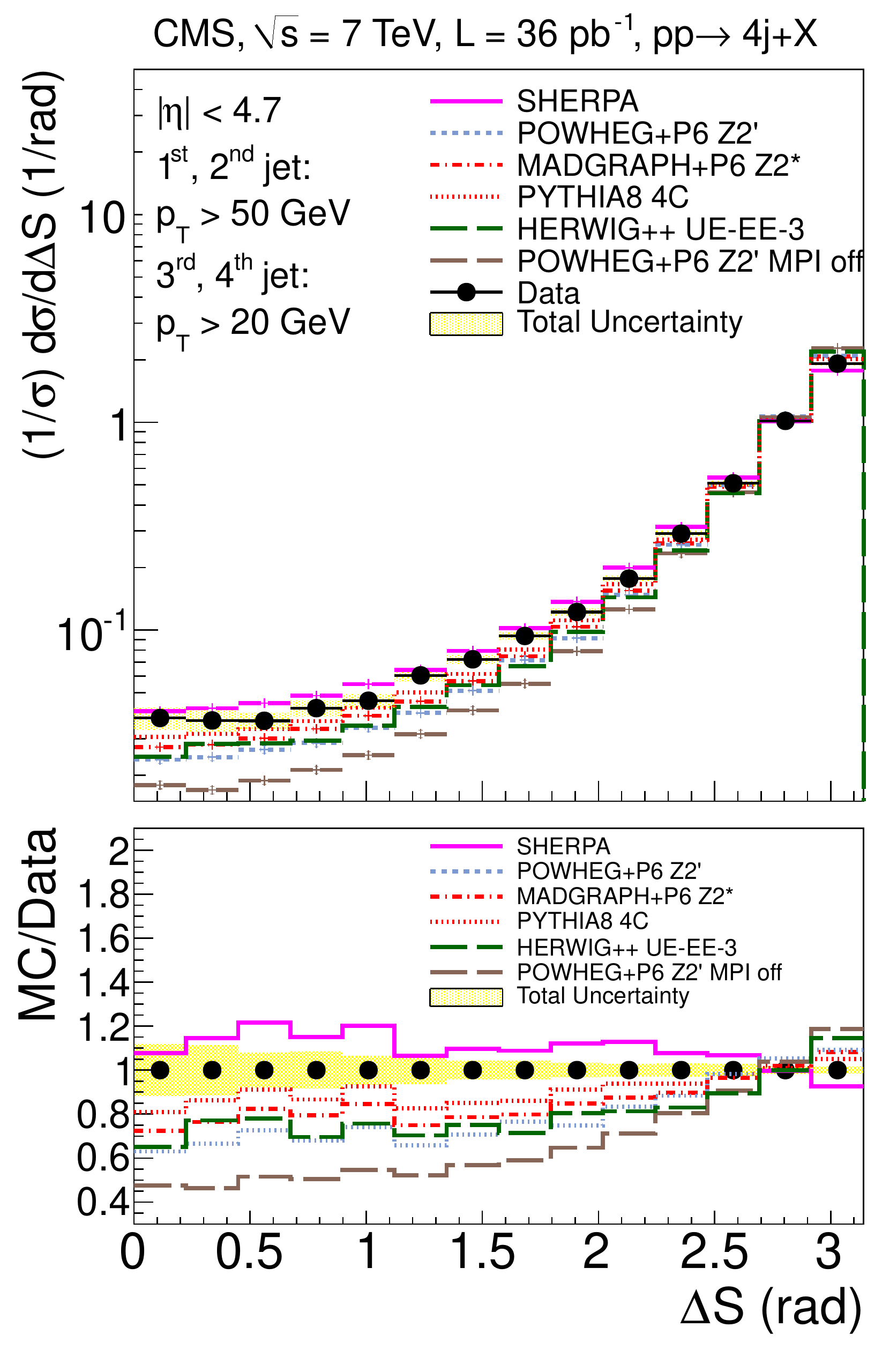}
 \caption{The differential cross sections for the production of exactly four jets,
          measured by the CMS collaboration, as a function of the jet transverse momenta 
	  \pt (left) and the azimuthal angle $\Delta\textrm{S}$ between the pair of the two 
	  leading and the pair of the two subleading jets(right), compared to predictions of 
	  \POWHEG, \MADGRAPH, \SHERPA, \PYTHIAE, and \HERWIGPP.}
 \label{fig:CMS4Jets}
 \end{center}
\end{figure}

Considering a final state of higher complexity,  the differential cross section for the production of exactly 
four jets is interesting to study. This measurement has been performed by the CMS collaboration as a function of the jet transverse momentum 
\pt\ and rapidity $|\textrm{y}|$, c.f.\ Ref.~\refcite{CMS_4_jets}.
Events with four jets can be produced via a single hard parton scattering (SPS) process, where two or more partons 
at high \pt are produced, with the initial- and final-state QCD radiation resulting in additional jets at lower \pt.
Multiparton interactions (MPI) can lead to the production of events, where more than one partonic interaction 
has occured in the same pp collision. In this case a pair of hard jets and a pair of softer jets can be produced 
via double parton scattering (DPS) leading to events with four jets.  
The SPS and DPS processes result in different distributions of angular correlations in observables such as
the azimuthal angle $\Delta\textrm{S}$ between the two dijet pairs.
The analysis is based on a data sample collected in 2010, with an integrated luminosity of $36\, \mathrm{pb^{-1}}$. 
Figure~\ref{fig:CMS4Jets} shows the cross sections for the production of exactly four jets as a function 
of the jet transverse momenta \pt (left), and the azimuthal angle $\Delta\textrm{S}$ between the two 
dijet pairs (right), compared to predictions of various MC generators, like \POWHEG\cite{powheg1,powheg2} , 
\MADGRAPH\cite{madgraph} , \SHERPA, \PYTHIAE\cite{pythia8} and \HERWIGPP\cite{herwigpp} . 
It is found that the models considered are able to describe the differential cross sections only in some 
regions of the phase space. Especially, the $\Delta\textrm{S}$ distribution is not described by any of 
the predictions and this may be taken as an indication for the need of including DPS in the investigated models.

\begin{figure}[hbtp]  
 \begin{center}
 \includegraphics[width=0.32\textwidth]{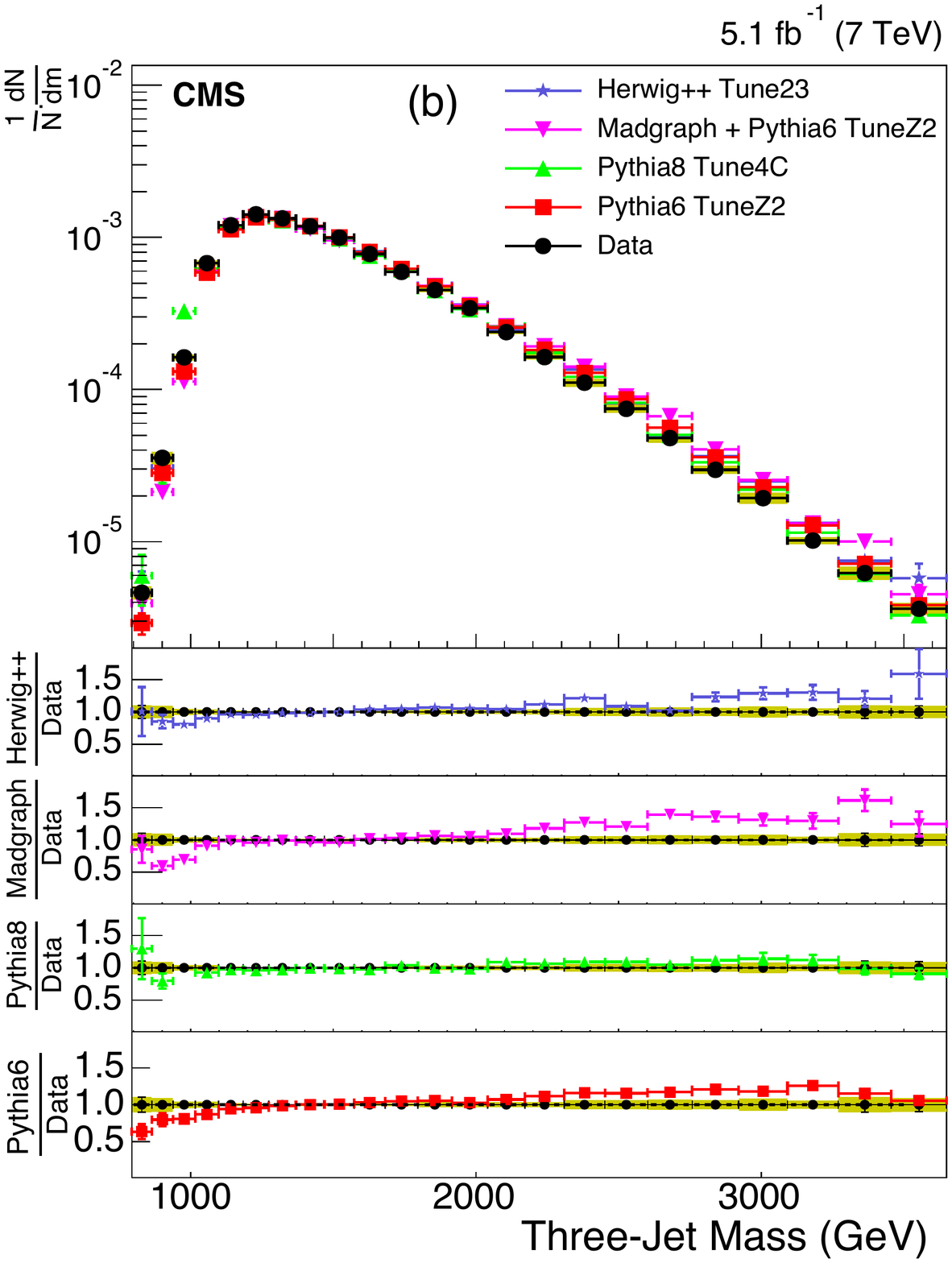}
 \includegraphics[width=0.32\textwidth]{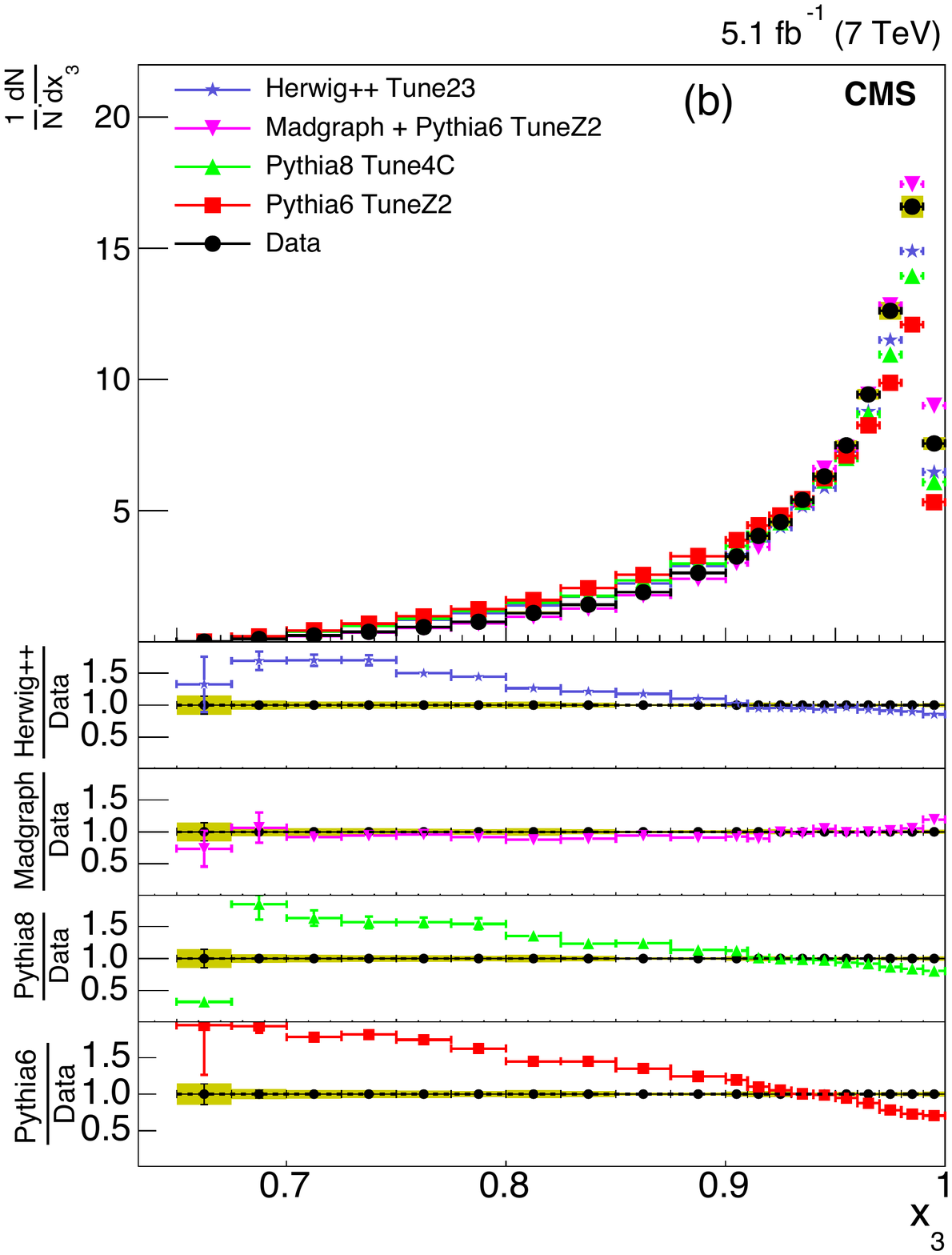}
 \includegraphics[width=0.32\textwidth]{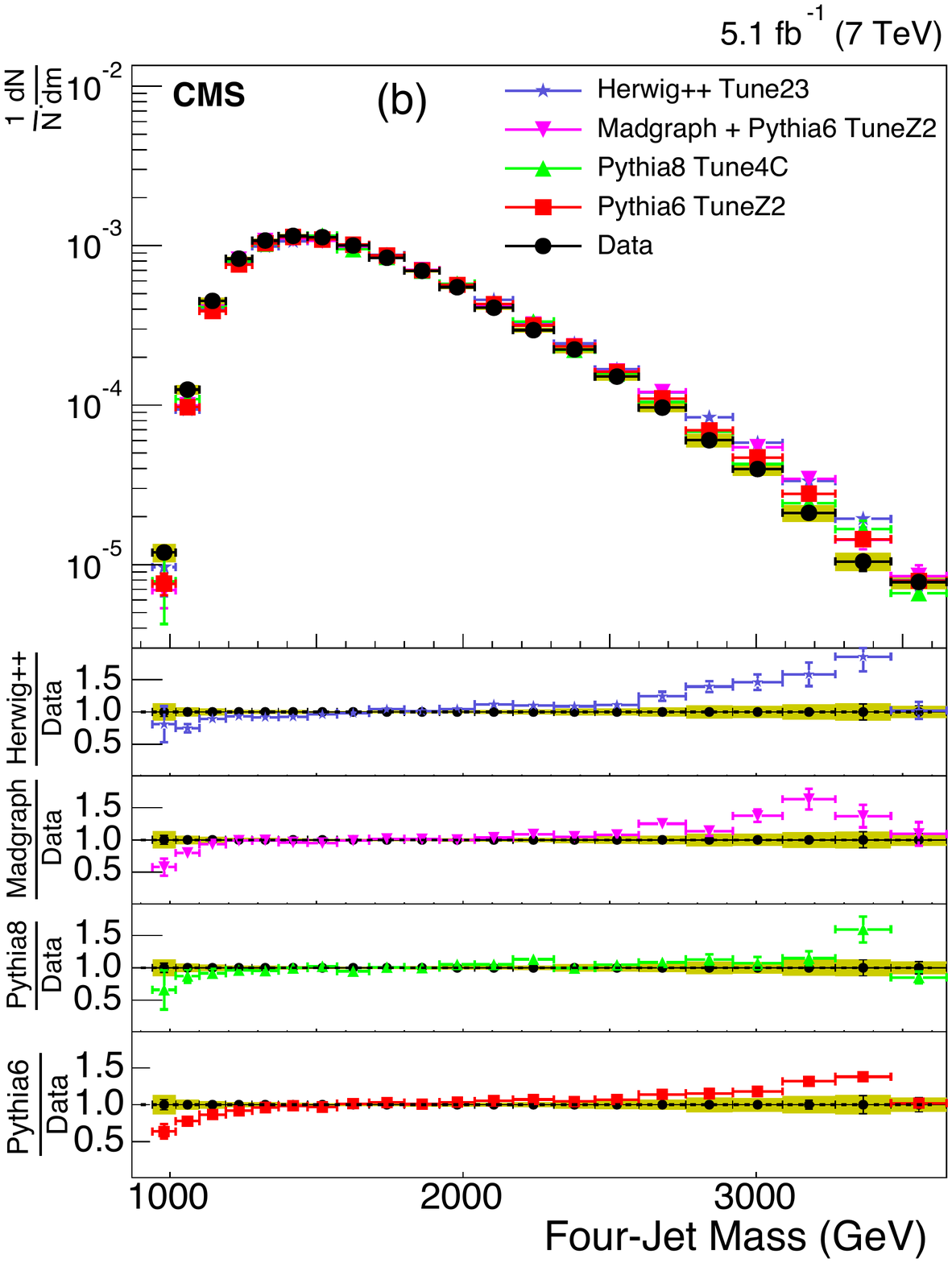}
 \caption{The three-jet mass (left), the scaled energy of the leading jet in the three-jet system (middle) 
          and the four-jet mass (right) measured by the CMS collaboration. 
	  Data are compared to predictions from four MC models: \PYTHIAS, \PYTHIAE, 
	  \MADGRAPH + \PYTHIAS, and \HERWIGPP.}
 \label{fig:CMSmultijets}
 \end{center}
\end{figure}

In Ref.~\refcite{CMS_multijet}, the CMS collaboration presents studies of inclusive topological 
distributions of three- and four-jet events using a data sample corresponding to an integrated 
luminosity of $5.1\, \mathrm{fb^{-1}}$ collected during 2011. 
These studies of three-jet events include the invariant mass of the system and the scaled energy 
distributions of the first and second jet. The scaled energy of the jet is defined as twice it's 
energy divided by the mass of the three-jet system. 
In the case of four-jet events, the four-jet mass and two event plane angles, namely the 
Nachtmann-Reiter\cite{nachtmannReiter} and the Bengtsson-Zerwas\cite{bengtssonZerwas} angle, are presented.
For example, in Fig.~\ref{fig:CMSmultijets} the three-jet mass, the scaled energy of the leading 
jet in the three-jet system, and the four-jet mass are presented. Data are compared to predictions 
of various LO MC models, with those from  \MADGRAPH + \PYTHIAS being closest to the data for most 
of the observables.
The multi-jet observables presented in this study are sensitive to higher-order processes and the
approximations used in their treatment. Furthermore, they provide a good test of various LO MC models 
widely used at the LHC.

\section{Ratios of jet cross sections}

The advantages of measuring ratios of jet cross sections are the reduction of several experimental
uncertainties, such as uncertainties in the jet energy scale and in the luminosity determination,
as well as of several theoretical systematic uncertainties related to the choice of the 
renormalisation and factorisation scales, $\mu_{r}$ and $\mu_{f}$, and to NP effects.
For these reasons ratios of jet cross sections provide a nice tool for testing pQCD, tuning MC models, 
constraining PDFs or determining the strong coupling constant.  

\begin{figure}[hbtp]  
 \begin{center}
 \includegraphics[width=0.48\textwidth]{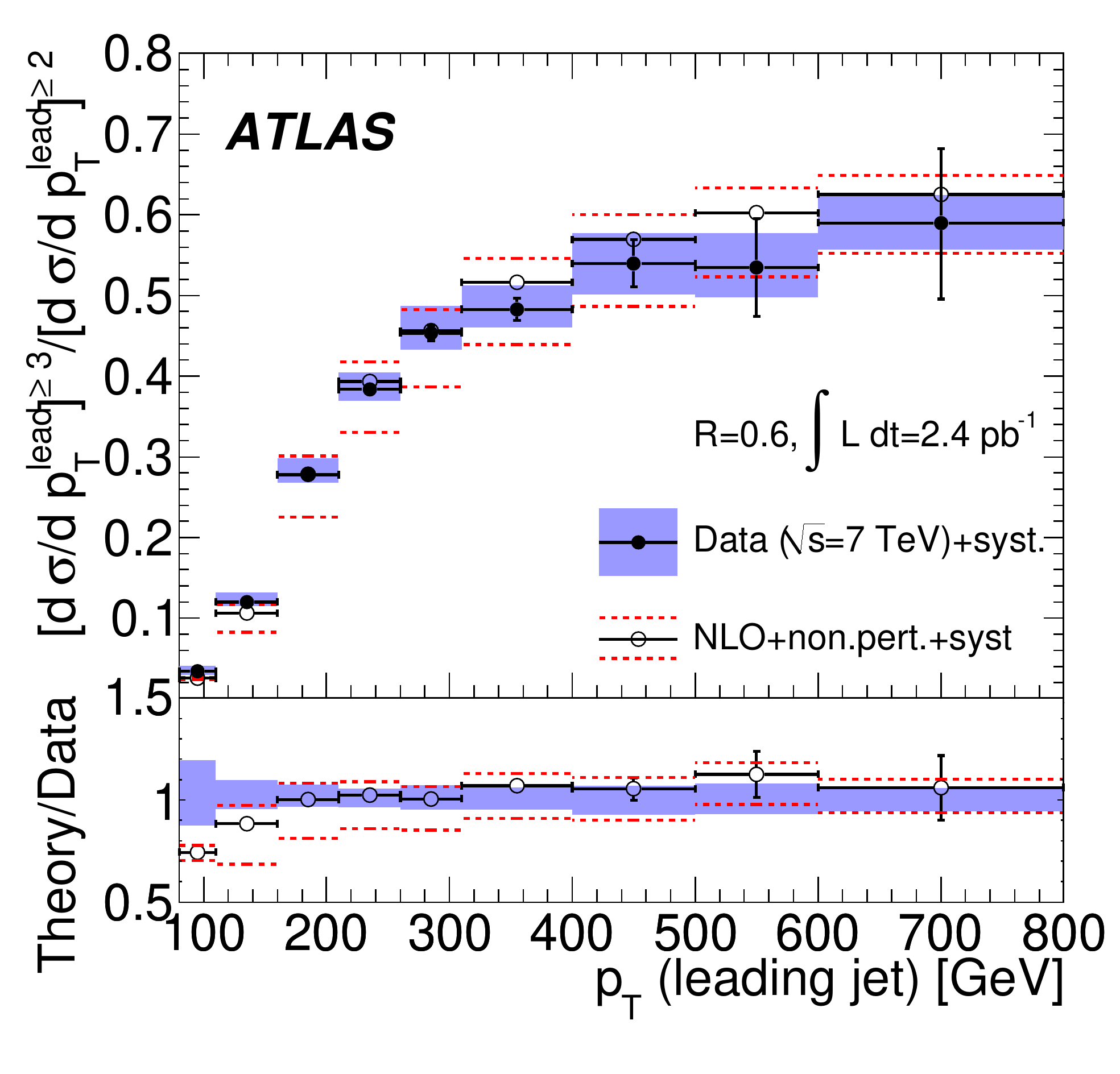}
 \includegraphics[width=0.48\textwidth]{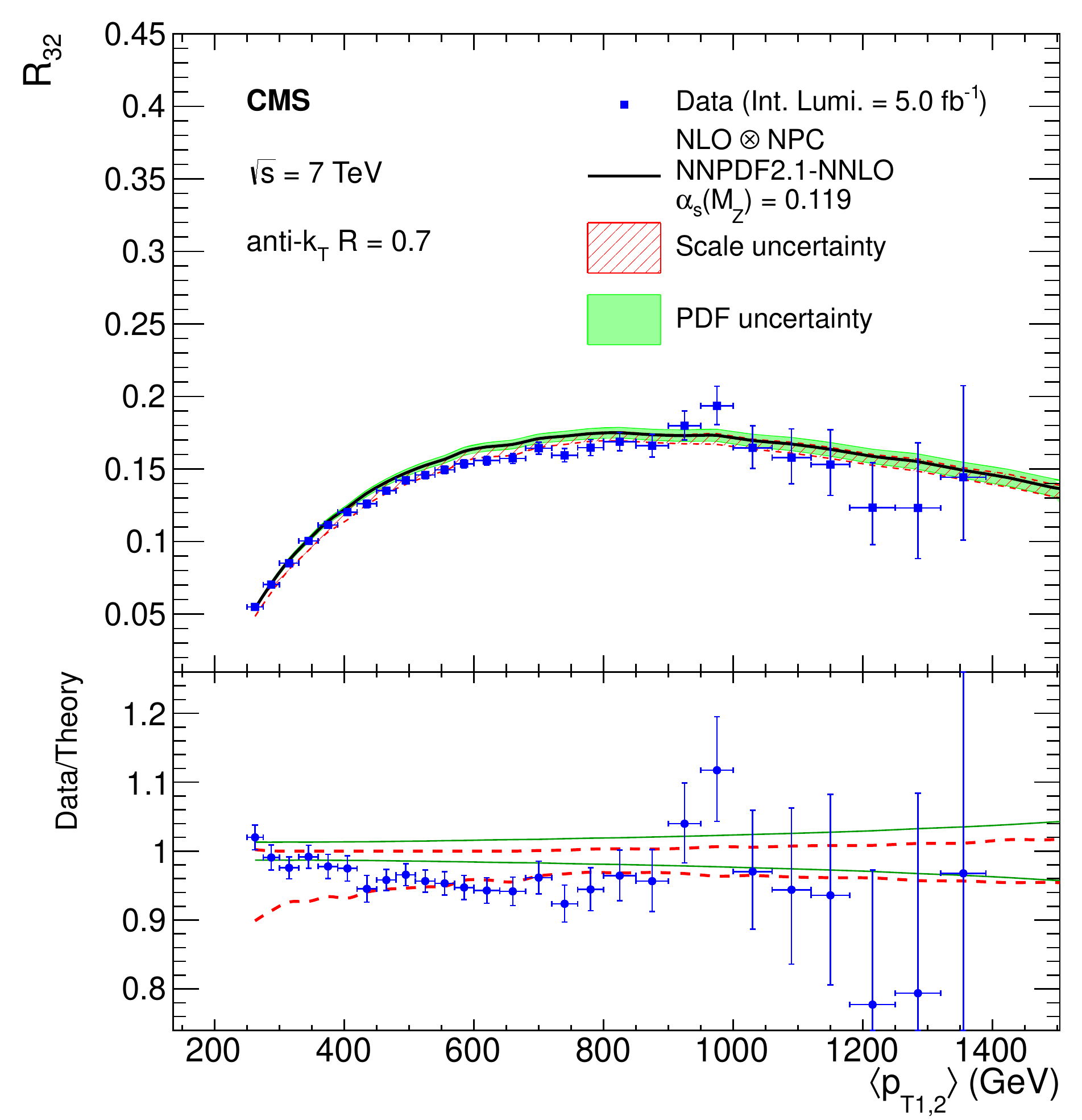}
 \caption{The ratio $R_{32}$ together with the NLO theoretical prediction,
          corrected for NP effects, measured by ATLAS (left) and CMS (right).
          In the bottom panels the ratio of data to 
	  the theoretical predictions is shown.}
 \label{fig:ATLAS_CMS_R32}
 \end{center}
\end{figure}

A measurement with particular sensitivity to limitations in the LO MC simulations and NLO pQCD 
calculations is the ratio $R_{32}$ of the inclusive 3-jet cross section to the inclusive 2-jet cross section. 
In Ref.~\refcite{ATLAS_multijets}, the ATLAS collaboration presents the measurement of $R_{32}$ 
as a function of the leading jet \pt and the total jet transverse momentum of the event, $H_\textrm{T}$. 
The results are compared to LO MC simulations, as well as to NLO theoretical predictions 
corrected for NP effects.
For example, Fig.~\ref{fig:ATLAS_CMS_R32} (left) shows the measured $R_{32}$ distribution together 
with the NLO prediction. 
Within uncertainties a nice agreement between data and theory is observed. 

The CMS collaboration has presented measurements of $R_{32}$ in two publications, see Refs.~\refcite{CMS_R32_Ht} 
and \refcite{CMS_R32_as}.
In the first publication, $R_{32}$ is measured versus $H_\textrm{T}$, using an integrated luminosity of 
$36\, \mathrm{pb^{-1}}$ collected during 2010. The measurement is used to test the validity of various 
MC generators such as \PYTHIAS, \PYTHIAE, \HERWIGPP, \MADGRAPH, and \ALPGEN, at multi-\TeV scales.
All MC generators describe the data within 20\%, with \MADGRAPH giving the best results.

In a second publication, CMS presented a measurement of $R_{32}$ as a function of the average transverse 
momentum of the two jets leading in \pt, using an integrated luminosity of $5\, \mathrm{fb^{-1}}$ collected during 2011. 
Data are compared to NLO theoretical predictions using various PDF sets.
For example, Fig.~\ref{fig:ATLAS_CMS_R32} (right) shows the measured ratio $R_{32}$  together 
with the NLO prediction using the NNPDF2.1 PDF set. 
Within uncertainties most PDF sets are able to describe the data. 
This $R_{32}$ measurement, due to its sensitivity to the strong coupling constant, has been used by the CMS collaboration 
for the first determination of $\alpha_{s}$ in the \TeV region, see Ref.~\refcite{Juan}.

\begin{figure}[hbtp]  
  \begin{center}
  \includegraphics[width=0.47\textwidth]{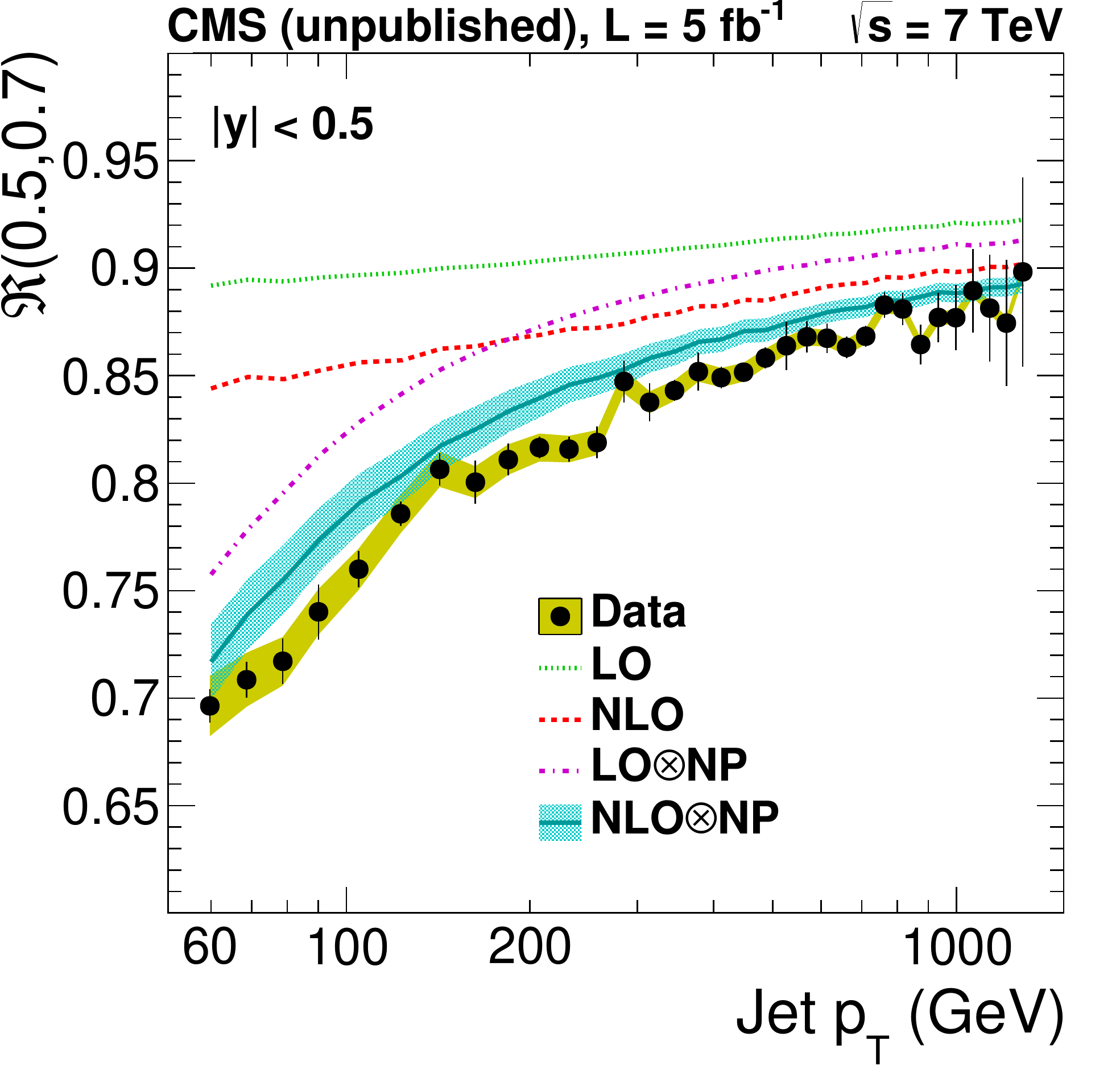}
  \includegraphics[width=0.47\textwidth]{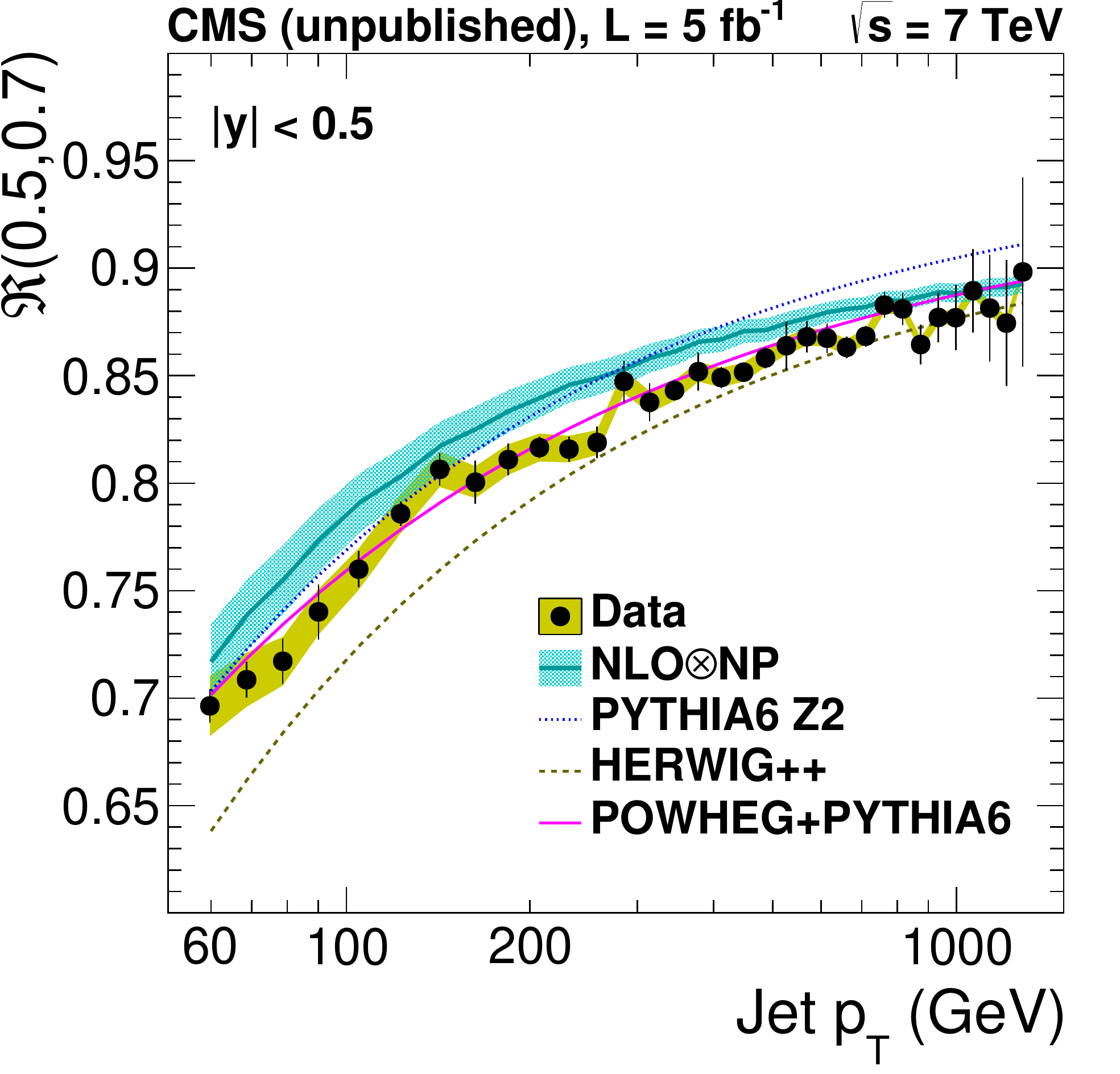}
  \caption{The jet radius ratio $\mathcal{R}(0.5, 0.7)$,  measured by the CMS collaboration, 
           in the central rapidity bin $|y| < 0.5$ 
           compared to LO and NLO predictions with and without NP corrections (left),  as well as to 
    NLO calculations corrected for NP effects ($\mathrm{NLO}\otimes\mathrm{NP}$) and to MC predictions (right). 
	   The bands represent the total correlated systematic uncertainty, while the error 
	   bars indicate the total uncorrelated uncertainty.}
  \label{fig:AK7AK5}
  \end{center}
 \end{figure}

A three-jet observable which is sensitive to the pattern of QCD radiation at NLO (including terms 
up to $\alpha_{s}^4$ ), is the ratio $\mathcal{R}(0.5, 0.7)$ of the inclusive jet cross sections using 
the \antikt clustering algorithm with two radius parameters, $R = 0.5$ and $0.7$.
The CMS collaboration has presented the measurement of $\mathcal{R}(0.5, 0.7)$ in Ref.~\refcite{CMS_ak7ak5}.
Figure \ref{fig:AK7AK5} shows $\mathcal{R}(0.5, 0.7)$ in the central rapidity bin $|y| < 0.5$,
compared to LO and NLO predictions with and without NP corrections (left),  as well as to 
NLO calculations corrected for NP effects and to MC predictions (right).
The study of this observable shows that models using LO (\PYTHIAS, \HERWIGPP) or NLO matrix element 
calculations matched to the parton showers (\POWHEG+\PYTHIAS), give a better description than the fixed order 
calculations corrected for NP effects. The best description for $\mathcal{R}(0.5, 0.7)$ is obtained by 
\POWHEG+\PYTHIAS. This demonstrates that jet radius dependent effects, measurable in data,
require pQCD predictions with at least one order higher than NLO or a combination of NLO
matrix elements and parton showers.

\section{Jet Shapes}

Jet shapes, the normalised transverse momentum flow as a function of the distance to the 
jet axis, provide information about the details of the parton-to-jet fragmentation process, 
see e.g.\ Ref.~\refcite{Jet_shapes_Ellis}. The ATLAS and CMS collaborations have provided the first measurements
of jet shapes in proton-proton collisions at the LHC, see Refs. \refcite{ATLAS_jet_shapes}
and \refcite{CMS_jet_shapes}.

Traditionally, the internal structure of a jet is studied in terms of the
differential and integrated jet shapes. The differential jet shape $\rho(r)$,
with $r=\sqrt{\Delta y^2 + \Delta \phi^2}$ being the distance to the jet axis, 
is defined as the average fraction of the transverse momentum contained inside 
an annulus of inner and outer radius $r-\delta r/2$ and $r+\delta r/2$ around the jet axis.
Alternatively, the integrated jet shape $\Psi(r)$ is defined as the average fraction 
of the transverse momentum of particles inside a cone of radius $r$ around the jet axis.

\begin{figure*}[hbtp]
 \begin{center}
   \includegraphics[width=0.55\linewidth]{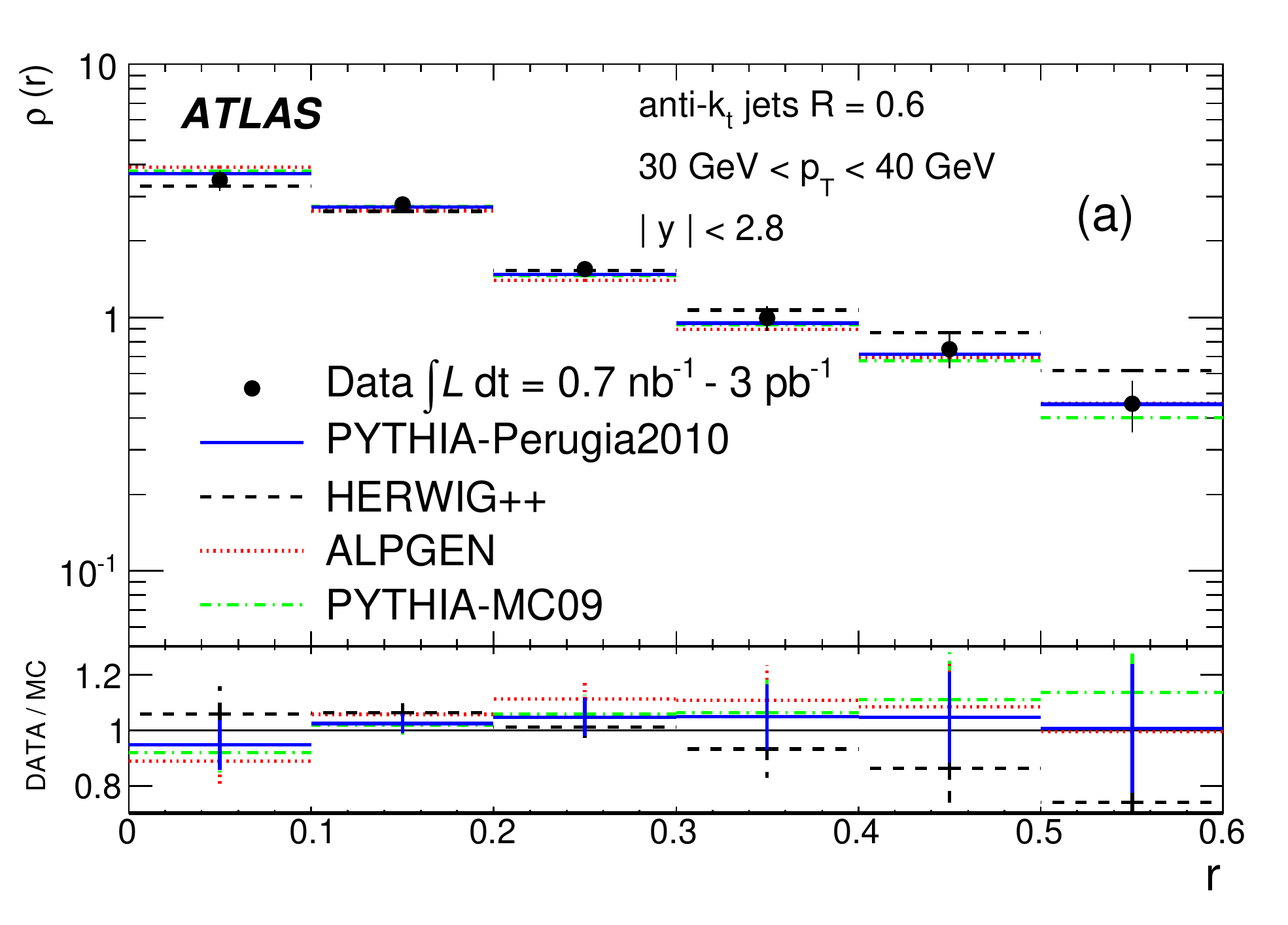}
   \includegraphics[width=0.42\linewidth]{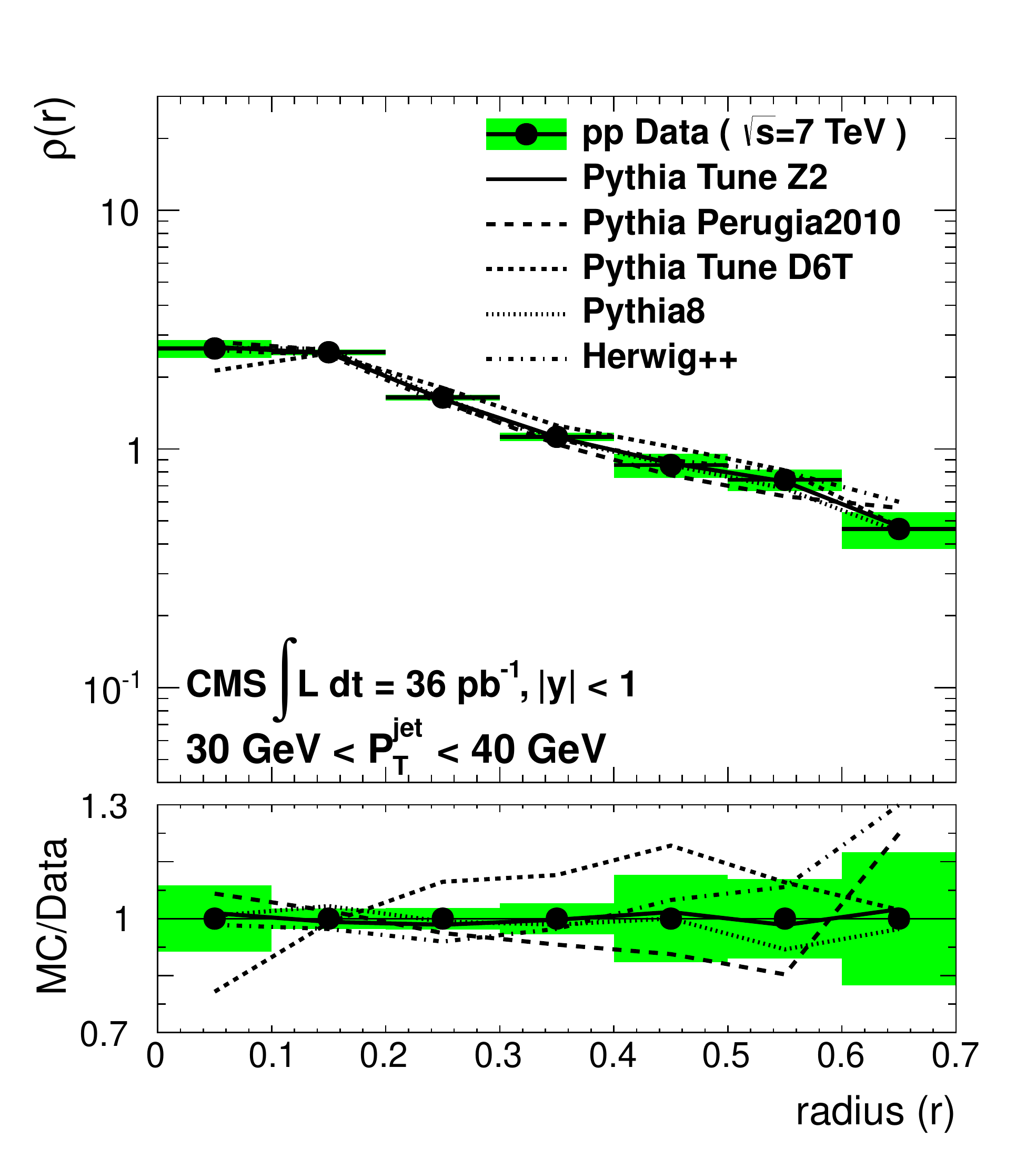}
   \includegraphics[width=0.55\linewidth]{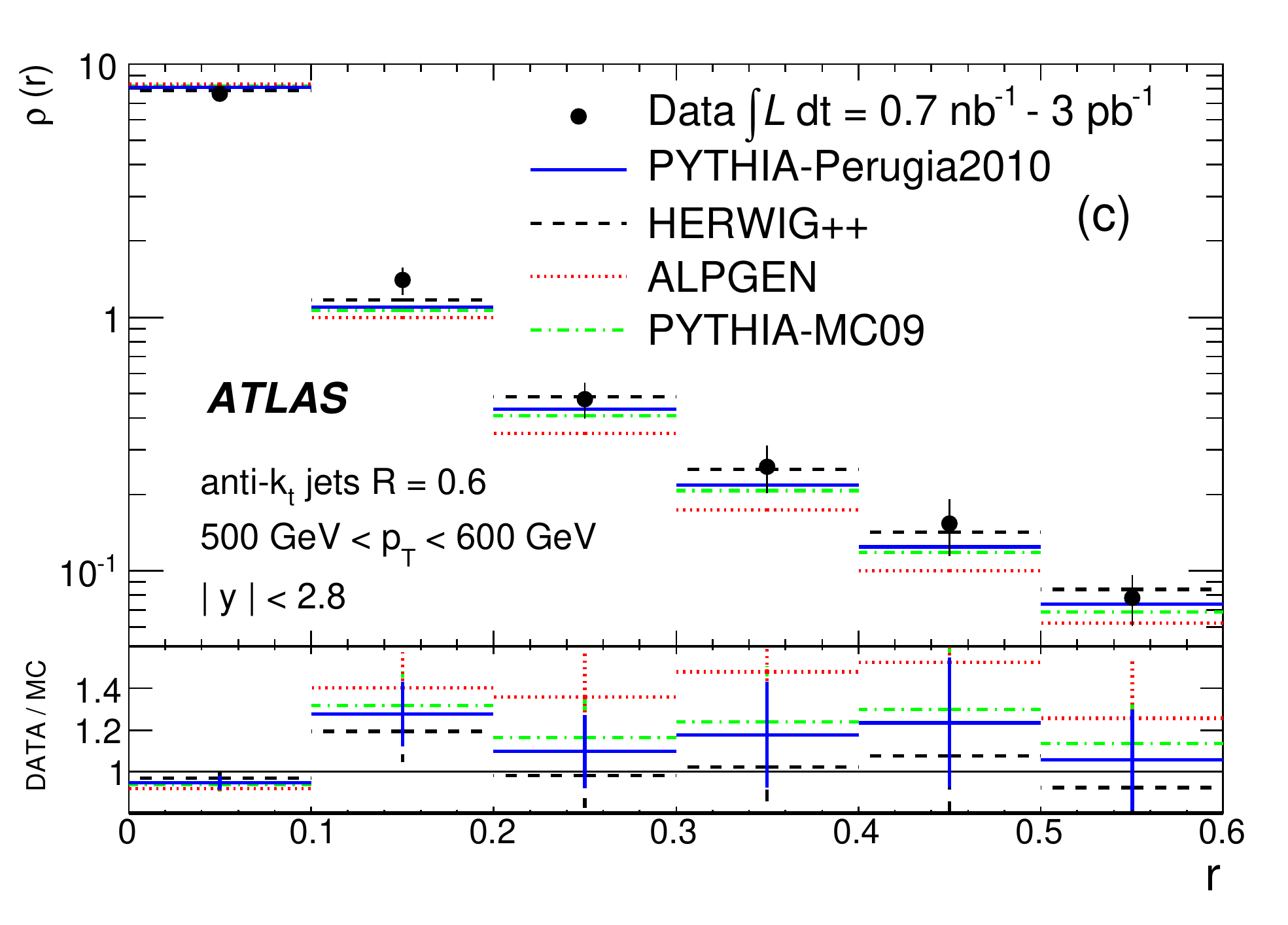}
   \includegraphics[width=0.42\linewidth]{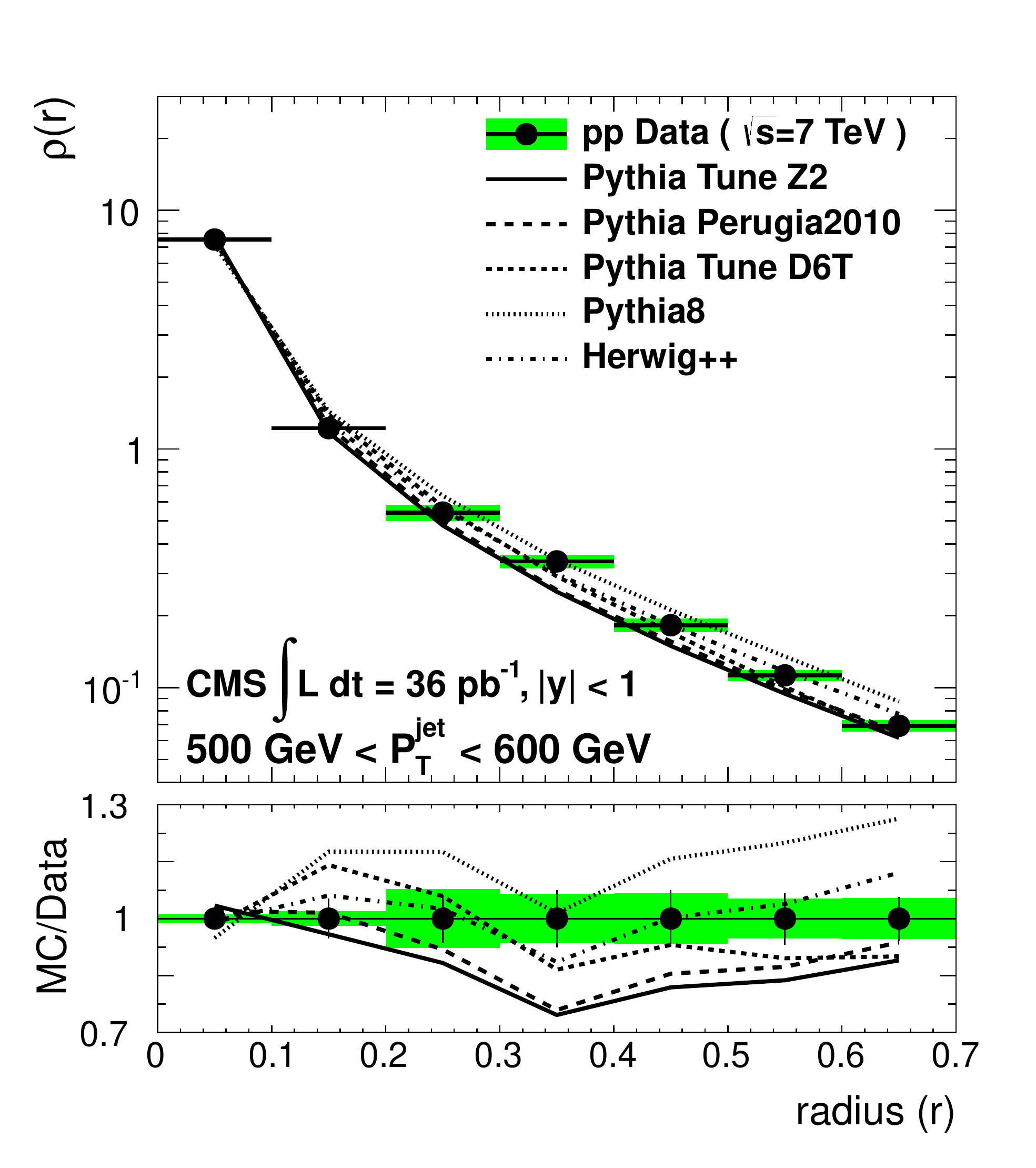}
   \caption{The measured differential jet shape $\rho(r)$ from ATLAS for $|y|\,<\,2.8$ (left) and 
            CMS for $|y|\,<\,1$ (right) and for two representative bins in jet \pt,
            $30\,\GeV\,<\,\pt\,<\,40\,\GeV$ (top) and
	    $500\,\GeV\,<\,\pt\,<\,600\,\GeV$ (bottom).
	    The ATLAS data are compared to predictions of PYTHIA-Perugia2010 (solid lines), 
	    HERWIG++ (dashed lines), ALPGEN interfaced with HERWIG and JIMMY (dotted lines), 
	    and PYTHIA-MC09 (dash-dotted lines).
	    The CMS data are compared to HERWIG++, PYTHIA8, and PYTHIA6 predictions with various tunes.}
   \label{fig:Jet_shapes_rho}
 \end{center}
\end{figure*}

\begin{figure*}[hbtp]
 \begin{center}
   \includegraphics[width=0.55\linewidth]{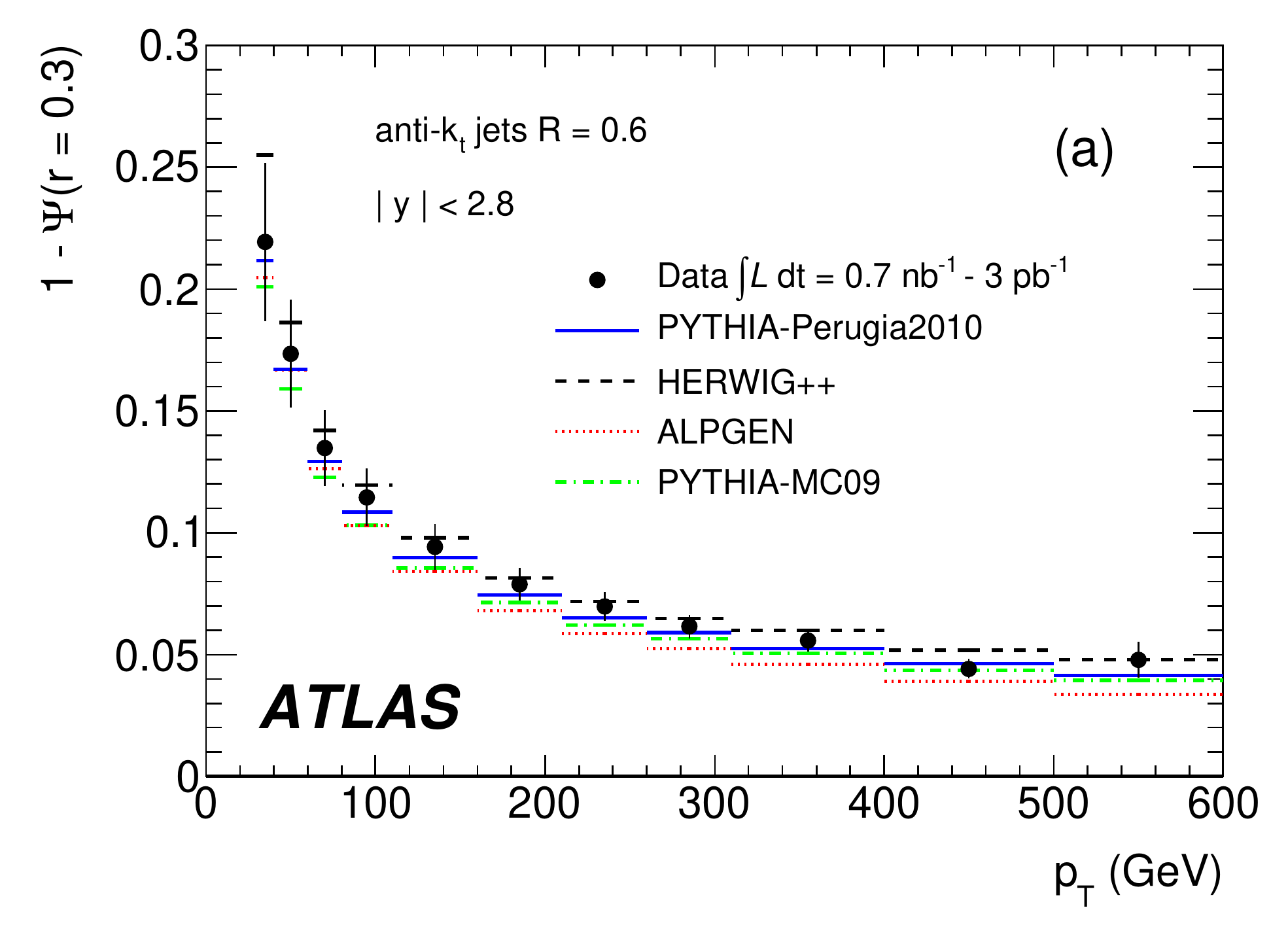}
   \includegraphics[width=0.42\linewidth]{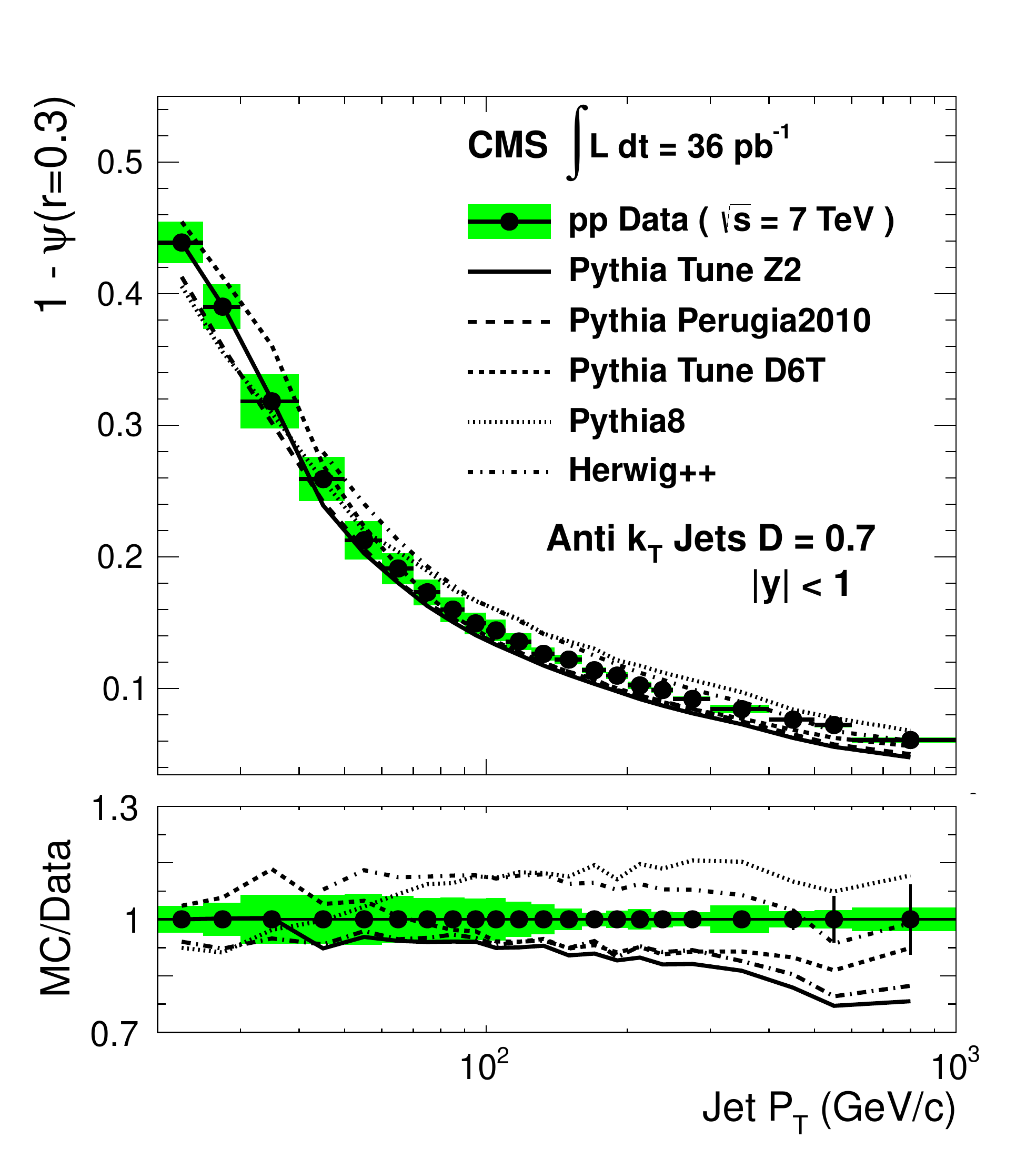}
   \caption{The integrated jet shape $1-\Psi(r=0.3)$ measured by ATLAS for $|y|\,<\,2.8$ (left) and by CMS for $|y|\,<\,1$ (right)
            as a function of jet $p_T$. The ATLAS data are compared to the predictions of PYTHIA-Perugia2010
            (solid lines), HERWIG++ (dashed lines), ALPGEN interfaced with HERWIG and JIMMY (dotted lines), and PYTHIA-MC09
            (dash-dotted lines). The CMS data are compared to HERWIG++, PYTHIA8, and PYTHIA6 predictions with various tunes.}
   \label{fig:Jet_shapes_psi}
 \end{center}
\end{figure*}

Figure~\ref{fig:Jet_shapes_rho} shows the differential jet shape $\rho(r)$ measured by 
ATLAS, for $|y|\,<\,2.8$ (left) and by CMS for $|y|\,<\,1$ (right), and for two
represenative bins in jet \pt, $30\,\GeV<\pt<40\,\GeV$ (top) 
and $500\,\GeV<\pt<600\,\GeV$ (bottom).
The peak at small $r$ indicates that the majority of the jet momentum is concentrated close 
to the jet axis. At high jet $\pt$ the peak is higher, indicating that jets are highly  
collimated with most of their $\pt$ close to the jet axis. About 95\% of the transverse 
momentum is contained within a cone of radius $r=0.3$.
This fraction decreases down to 80\% at low jet $\pt$, where jets become wider. 
This is also demonstrated in Fig.~\ref{fig:Jet_shapes_psi}, where the measured integrated 
jet shape $1-\Psi(r=0.3)$ from ATLAS (left) and CMS (right) is shown as a function of jet \pt.
Data from ATLAS and CMS are compared to predictions of various MC generators such as
PYTHIA6, PYTHIA8, HERWIG++, and ALPGEN in various tunes, thus giving interesting sensitivity
to a variety of perturbative and non-perturbative effects.

\begin{figure*}[hbtp]
 \begin{center}
   \includegraphics[width=0.40\linewidth]{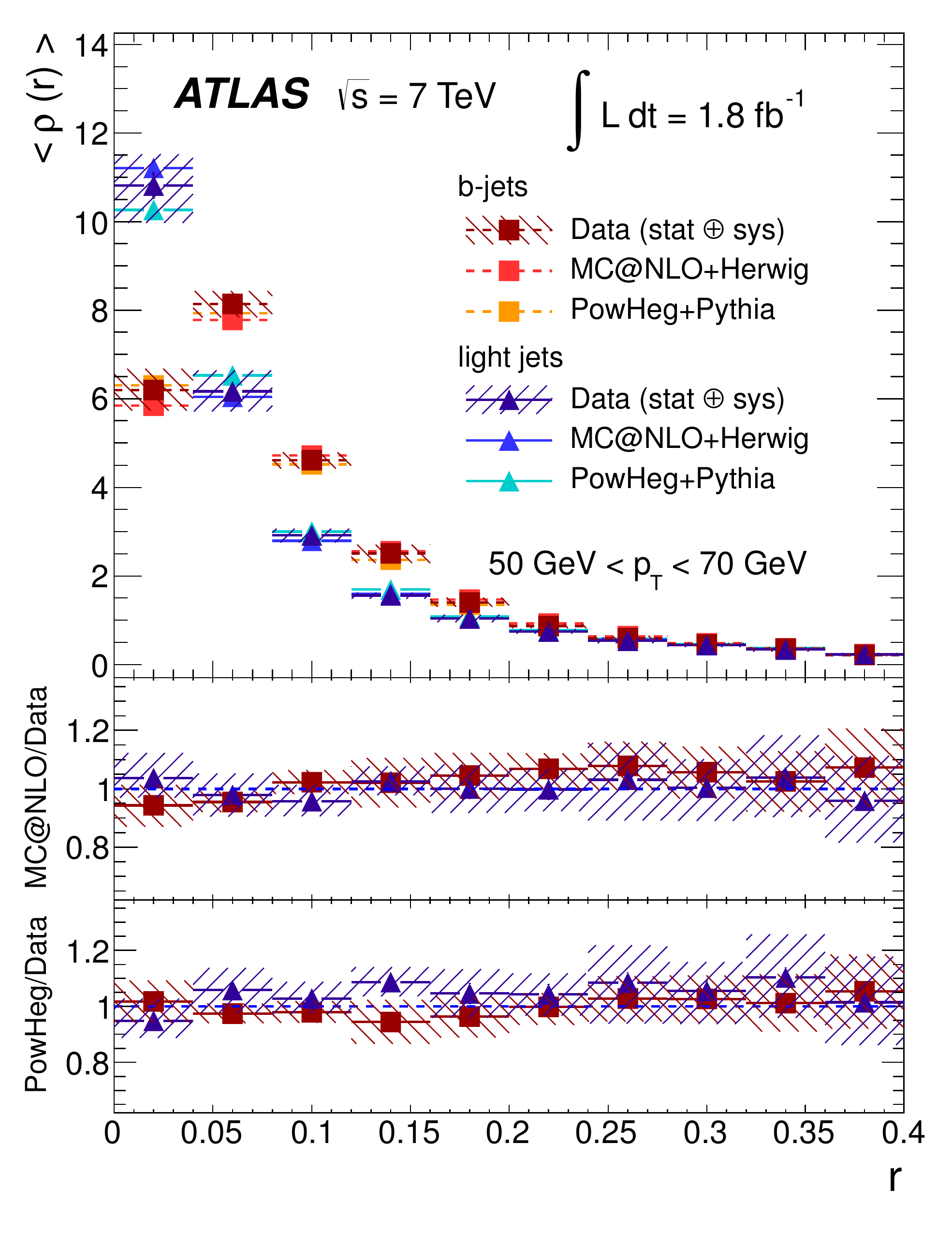}
   \includegraphics[width=0.40\linewidth]{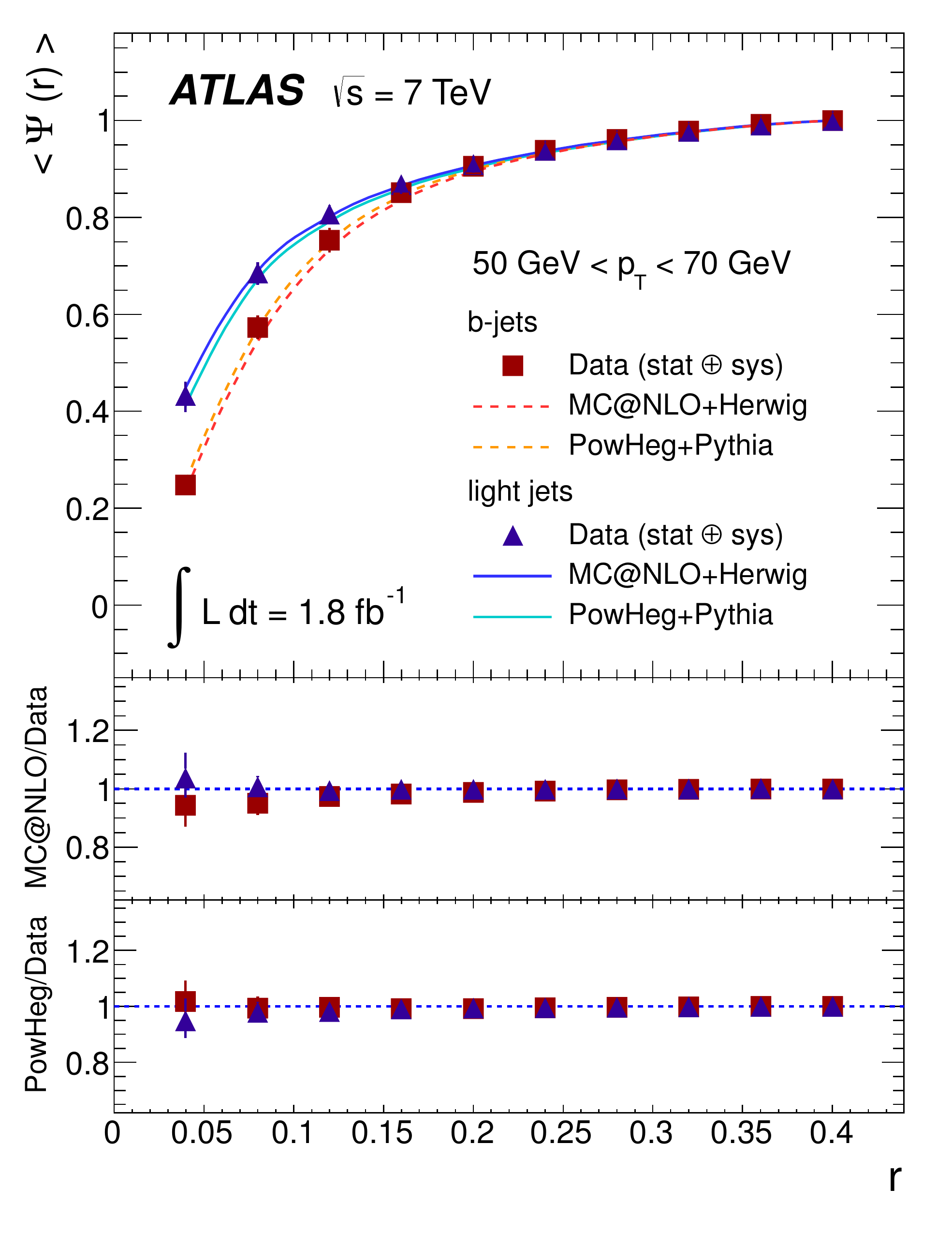}
   \caption{The differential ($\langle\rho(r)\rangle$, left) and the integrated 
            ($\langle\Psi(r)\rangle$, right) jet shapes, measured by ATLAS, as a function 
	    of the radius $r$ for b-jets (squares) and light jets (triangles) in the interval 
	    $100\,\GeV<\pt<150\,\GeV$. 
	    Data are compared to the $\mathrm{MC@NLO}$+HERWIG and POWHEG+PYTHIA event generators.}
   \label{fig:Jet_shapes_ttbar}
 \end{center}
\end{figure*}

In Ref.~\refcite{ATLAS_jet_shapes_ttbar} ATLAS also presented the first measurement of b-jet 
shapes in top pair events. This measurement allows the study of the differences between 
the b-quark and light-quark jets, using the differential and integrated jet shapes. 
Figure~\ref{fig:Jet_shapes_ttbar} shows the differential ($\langle\rho(r)\rangle$, left) 
and the integrated ($\langle\Psi(r)\rangle$, right) jet shapes, as a function 
of the radius $r$ for b-jets (squares) and light jets (triangles) in a representative bin of jet 
\pt, $100\,\GeV<\pt<150\,\GeV$.
The analysis shows that b-jets are broader than light jets, with the cores of light jets having a larger 
energy density than those of b-jets. The measurement of the integrated  
jet shape, $\langle\Psi(r)\rangle$, shows that for low values of $r$ it is possible to distinguish b-jets from light jets.
The MC event generators $\mathrm{MC@NLO}$\cite{mcNLO}+HERWIG and POWHEG+PYTHIA reproduce well the jet 
shapes for both light and b-jets.

\section{Jet Event Shapes}

Event shape variables are geometric properties of the energy flow in hadronic final states.
They are sensitive to QCD radiation, as gluon emission changes the shape of the energy flow,
providing an indirect probe of multi-jet topologies.

ATLAS (in Ref.~\refcite{ATLAS_event_shapes}) and CMS (in Refs.~\refcite{CMS_event_shapes1} 
and \refcite{CMS_event_shapes2}) presented the first event shapes measurements at LHC, 
for testing the validity of various MC generators.   
These include: the event thrust $\tau_\bot$ and its minor component $T_{m,\bot}$ ;
the sphericity $S$ and its transverse component $S_\bot$; the aplanarity A;
the jet broadening $B_\mathrm{tot}$; 
the total jet mass $\rho_\mathrm{tot}$ and its transverse component $\rho_\mathrm{tot}^{T}$;
and the third-jet resolution parameter $\Upsilon_{23}$. 
Phenomenological discussions and definitions of event shapes at hadron colliders can be 
found in Refs.~\refcite{Event_shapes_phen1} and \refcite{Event_shapes_phen2}.

\begin{figure*}[hbtp]
 \begin{center}
   \includegraphics[trim = 80mm 0mm 0mm 0mm, clip, width=0.45\linewidth]{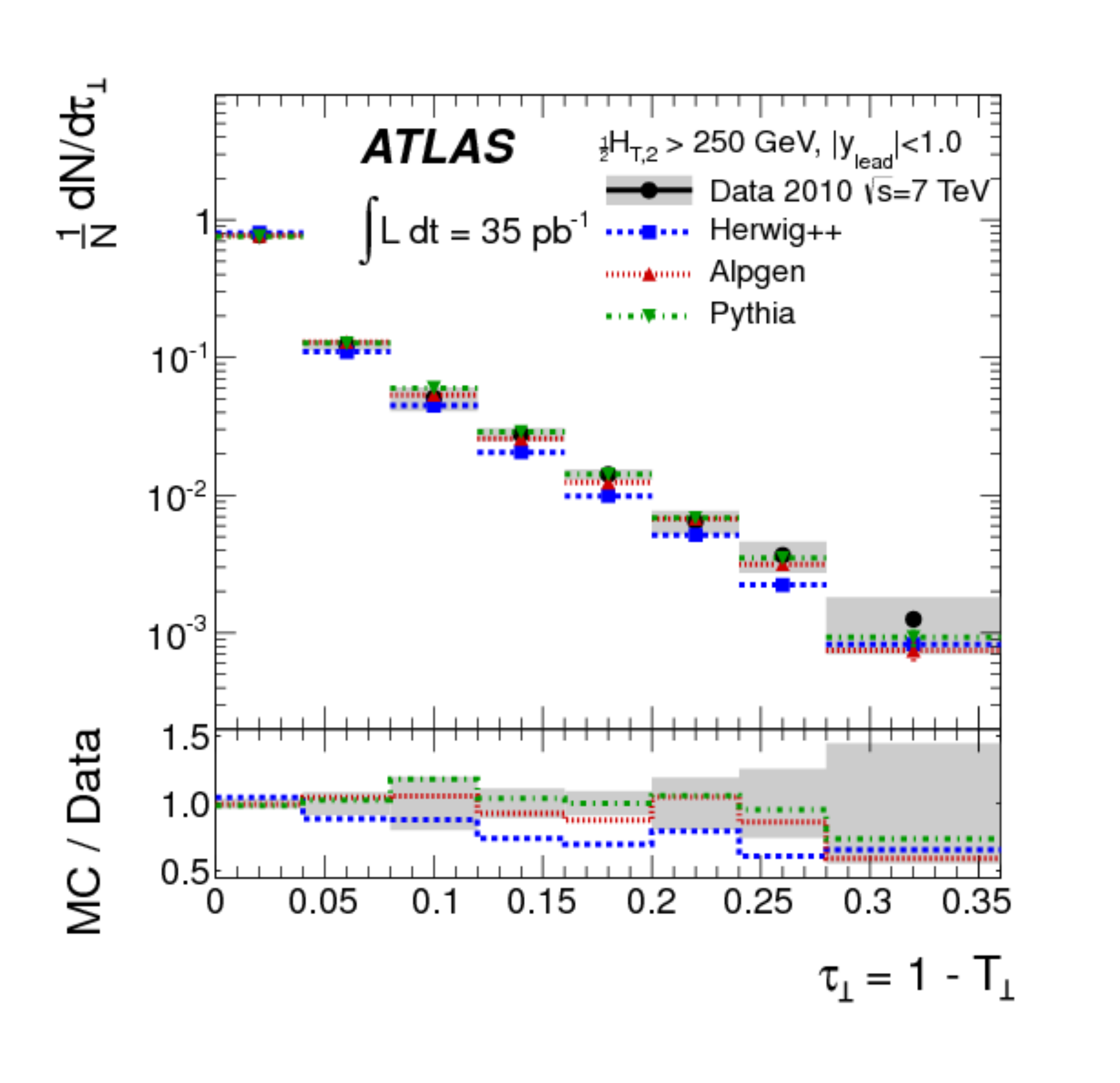}
   \includegraphics[trim = 80mm 0mm 0mm 0mm, clip, width=0.45\linewidth]{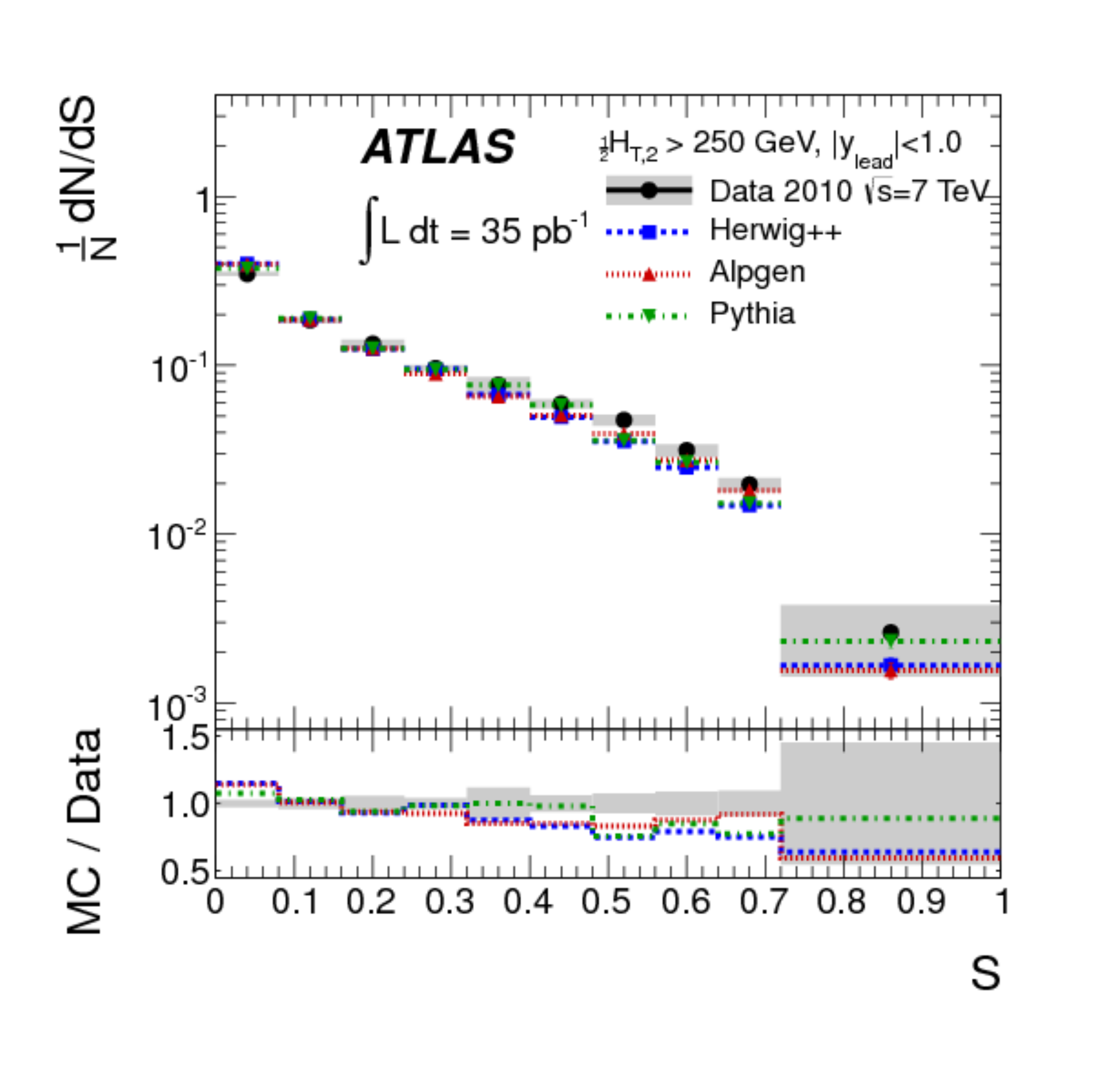}
   \caption{The event thrust (left) and the sphericity (right) measured by the ATLAS collaboration. 
          The results are compared to different MC simulations.}
   \label{fig:ATLAS_event_shapes}
 \end{center}
\end{figure*}

\begin{figure*}[hbtp]
 \begin{center}
   \includegraphics[trim = 0mm 0mm 0mm 173.5mm, clip, width=.84\linewidth]{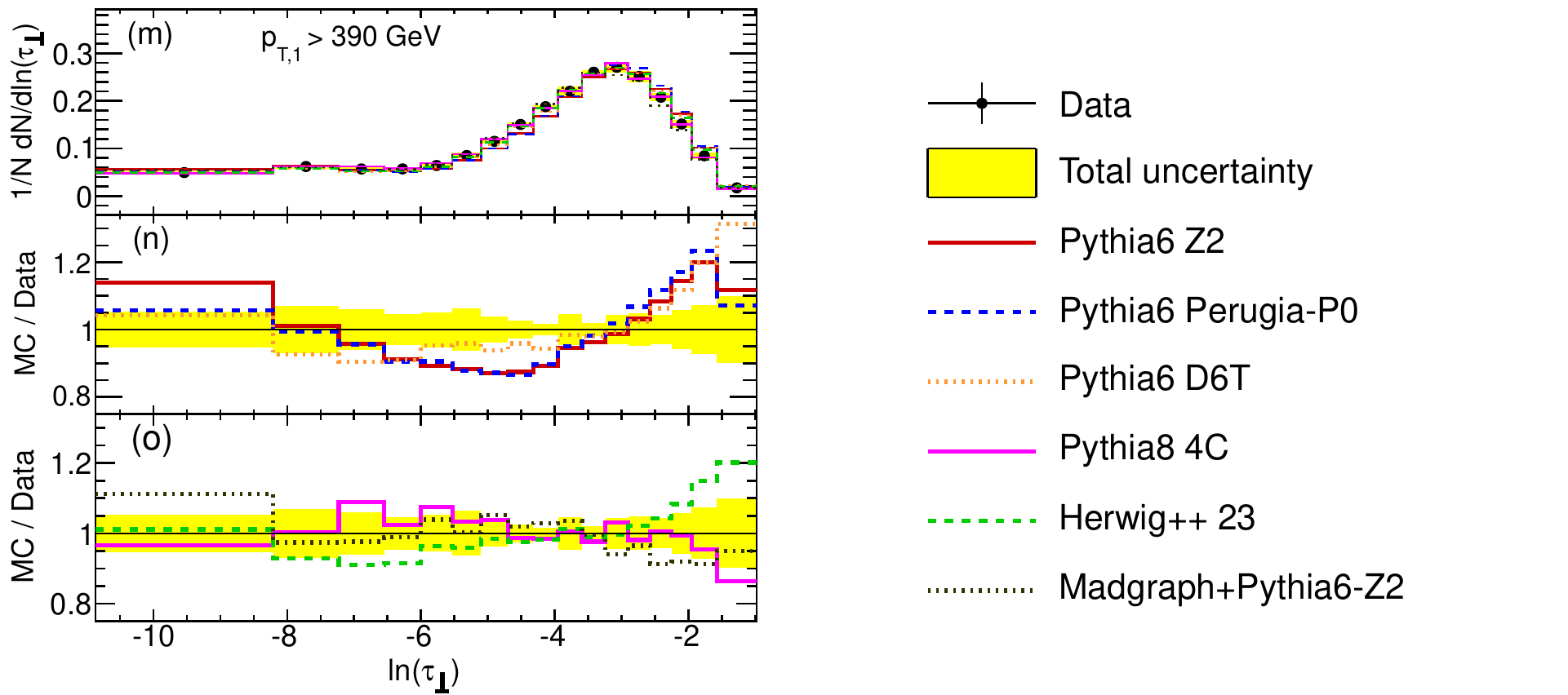}
   \includegraphics[trim = 0mm 0mm 99mm 173.5mm, clip, width=0.42\linewidth]{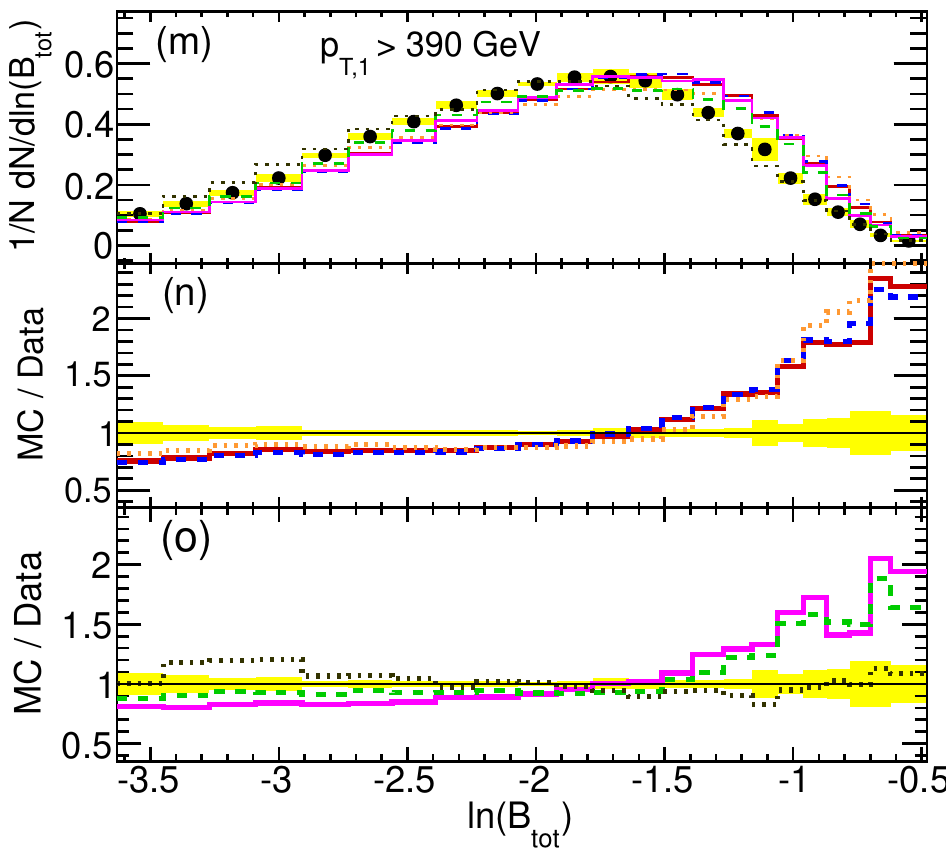}
   \includegraphics[trim = 0mm 0mm 99mm 173.5mm, clip, width=0.42\linewidth]{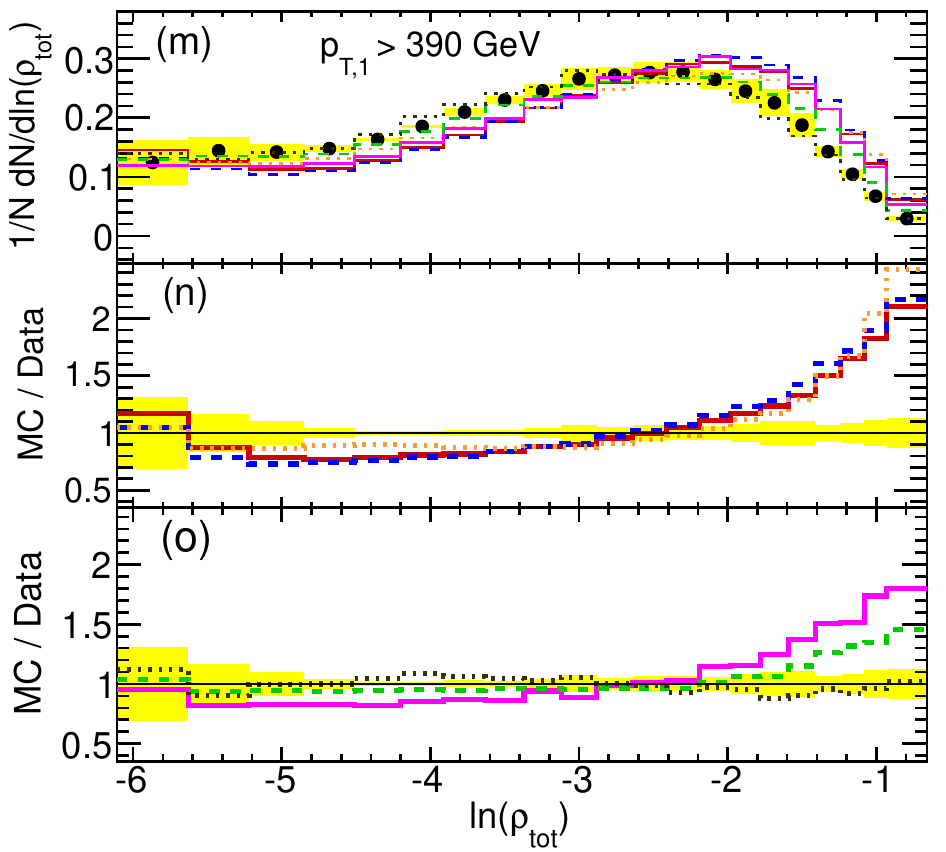}
   \caption{The event thrust (top left), the jet broadening (bottom left) and the total jet mass (bottom right) 
            distributions measured by the CMS collaboration. The results are compared to different MC simulations.}
   \label{fig:CMS_event_shapes}
 \end{center}
\end{figure*}

Figure~\ref{fig:ATLAS_event_shapes} shows the event thrust (left) and the sphericity (right), 
measured by the ATLAS collaboration, using an integrated luminosity of $35\, \mathrm{pb^{-1}}$ collected during 2010.
Data are compared to predictions of \HERWIGPP, \ALPGEN and \PYTHIA MC simulations, showing reasonable agreement.
In Fig.~\ref{fig:CMS_event_shapes}, the measurements of the event thrust (top left), the jet broadening 
(bottom left) and the total jet mass (bottom right) distributions, measured by the CMS collaboration, are presented.
This analysis uses an integrated luminosity of $5\, \mathrm{fb^{-1}}$ collected during 2011.
Data are compared to the \PYTHIAS, \PYTHIAE, \HERWIGPP, and \MADGRAPH MC generators.
For the event thrust, all generators show an overall agreement with data within 10\%, with \PYTHIAE and 
\HERWIGPP exhibiting a better agreement than the others. For the jet broadening and the total jet mass 
the agreement of the various MC generators with data is poor, except for \MADGRAPH, which 
provides a good description of the measurements.

\section{Conclusions}

A review has been presented on recent measurements of jet-related observables at the LHC,
such as multi-jet rates and cross sections, ratios of jet cross sections, jet shapes and event shape 
observables.
The measurements are based on the data collected during the first two years of the
LHC operation in proton-proton collisions at the center-of-mass energy of 7 TeV. 

These observables are compared to predictions of various widely used LO and some recent 
higher-order MC generators, providing valuable information regarding their usage at LHC energies.  
Furthermore, data are compared to predictions of NLO calculations, thus providing 
a test of pQCD in a previously unexplored energy region and constraining our knowledge 
of the Standard Model and its phenomenology.
QCD is the main background in many searches of new physics and in this sense the overall 
good agreement between data and theoretical predictions provides a solid foundation for searches 
beyond the Standard Model. 

In the coming years the continuation of the analyses, by ATLAS and CMS, on the 8 TeV 
data will provide more information on these jet-observables, completing the QCD physics program  
based on the data of the first three-years operation period of LHC.
And of course, even more data are anticipated in the near future, from the 13 TeV LHC run.

%%% \section*{Acknowledgments}

%%% This section should come before the References. Dedications and funding
%%% information may also be included here.

%\section*{References}

%\begin{thebibliography}{000} %for 3 digits
%\begin{thebibliography}{00}  %for 2 digits


\begin{thebibliography}{0}    %for 1 digit

%% ATLAS & CMS detectors
\bibitem{ATLAS_detector} ATLAS Collab. (G. Aad {\it et al}.),
{\it JINST\/} {\bf 3}, S08003 (2008).

\bibitem{CMS_detector} CMS Collab. (S. Chatrchyan {\it et al}.),
{\it JINST\/} {\bf 3}, S08004 (2008).


%% Inclusive jets-dijets
\bibitem{CMS_incl} CMS Collab. (S. Chatrchyan {\it et al}.),
{\it Phys Rev. Lett.\/} {\bf 107}, 132001 (2011).

\bibitem{CMS_dijet} CMS Collab. (S. Chatrchyan {\it et al}.),
{\it Phys. Lett. B\/} {\bf 700}, 187 (2011).

\bibitem{ATLAS_inclDijet2011} ATLAS Collab. (G. Aad {\it et al}.),
{\it Eur. Phys. J. C\/} {\bf 71}, 1512 (2011).

\bibitem{ATLAS_inclDijet2012} ATLAS Collab. (G. Aad {\it et al}.),
{\it Phys. Rev. D\/} {\bf 86}, 014022 (2012).

\bibitem{CMS_inclDijet} CMS Collab. (S. Chatrchyan {\it et al}.),
{\it Phys. Rev. D\/} {\bf 87}, 112002 (2013).

\bibitem{ATLAS_276} ATLAS Collab. (G. Aad {\it et al}.),
{\it EPJC\/} {\bf 73}, 2509 (2013).

\bibitem{ATLAS_incl7TeV} ATLAS Collab. (G. Aad {\it et al}.),
{\it JHEP\/} {\bf 02}, 153 (2015).

\bibitem{Francavilla} 
P. Francavilla, {\it Measurements of inclusive jet and dijet cross sections at the LHC}, article
in this volume of reviews.


%% Rates & cross sections

%% ATLAS multijets
\bibitem{ATLAS_multijets} ATLAS Collab. (G. Aad {\it et al}.),
{\it Eur. Phys. J. C\/} {\bf 71}, 1763 (2011).

\bibitem{pythia6} T. Sjostrand, S. Mrenna, and P. Z. Skands, 
{\it JHEP\/} {\bf 05}, 026 (2006).

\bibitem{alpgen} M.L. Mangano, 
{\it JHEP\/} {\bf 07}, 001 (2003).

\bibitem{sherpa} T. Gleisberg {\it et al}, 
{\it JHEP\/} {\bf 02}, 007 (2009).


%% ATLAS 3jet cross sections
\bibitem{ATLAS_3jetCrossSections} ATLAS Collab. (G. Aad {\it et al}.),
{\it Eur. Phys. J. C\/} {\bf 75}, 228 (2015)

\bibitem{ct10a} H.-L. Lai {\it et al}.,
{\it Phys. Rev. D\/} {\bf 82},074024  (2010).

\bibitem{ct10b} J. Gao {\it et al}.,
{\it Phys. Rev. D\/} {\bf 89}, 033009 (2014).

\bibitem{mstw2008a} A.D. Martin, W.J. Stirling, R.S. Thorne and G. Watt,
{\it Eur. Phys. J. C\/} {\bf 63}, 189 (2009).

\bibitem{mstw2008b} A.D. Martin, W.J. Stirling, R.S. Thorne and G. Watt,
{\it Eur. Phys. J. C\/} {\bf 64}, 653 (2009).

\bibitem{gjr08}  M. Gluck, P. Jimenez-Delgado, E. Reya, and C. Schuck
{\it Phys. Lett. B\/} {\bf 664}, 133 (2008).



%% 3Jet mass
\bibitem{CMS_3jetmass} CMS Collab. (S. Chatrchyan {\it et al}.),
{\it Eur. Phys. J. C\/} {\bf 75}, 288 (2015)

\bibitem{nnpdf1} R.D. Ball {\it et al}.,
{\it Nucl. Phys. B\/} {\bf 838}, 136 (2010).

\bibitem{nnpdf2} R.D. Ball {\it et al}.,
{\it Nucl. Phys. B\/} {\bf 849}, 296 (2011).

\bibitem{herapdf} H1 and ZEUS Collab. (F.D. Aaron {\it et al}.),
{\it JHEP\/} {\bf 01}, 109 (2010).

\bibitem{abm11} S. Alekhin, J. Blümlein and S. Moch,
{\it Phys. Rev. D\/} {\bf 86}, 054009 (2012).


%% Cross reference 
\bibitem{Juan} J. Rojo, {\it Constraints on parton distributions and the strong coupling from LHC jet data}, article
in this volume of reviews.



%%4 jets cross sections
\bibitem{CMS_4_jets} CMS Collab. (S. Chatrchyan {\it et al}.),
{\it Phys. Rev. D\/} {\bf 89}, 092010 (2014).

\bibitem{powheg1} S. Frixione, P. Nason, and C. Oleari, 
{\it JHEP\/} {\bf 11}, 070 (2007).

\bibitem{powheg2} S. Alioli, P. Nason, C. Oleari, and E. Re, 
{\it JHEP\/} {\bf 06}, 043 (2010).

\bibitem{madgraph} J. Alwall {\it et al}, 
{\it JHEP\/} {\bf 06}, 128 (2011).

\bibitem{pythia8} T. Sjostrand, S. Mrenna, and P. Z. Skands, 
{\it Phys. Commun.\/} {\bf 178}, 852 (2008).

\bibitem{herwigpp} M. Bahr {\it et al}, 
{\it Eur. Phys. J. C\/} {\bf 58}, 639 (2008).


%% nmultijet 
\bibitem{CMS_multijet} CMS Collab. (S. Chatrchyan {\it et al}.),
{\it Eur. Phys. J. C\/} {\bf 75}, 302 (2015)

\bibitem{nachtmannReiter} O. Nachtmann and A. Reiter, 
{\it Z. Phys. C\/} {\bf 16}, 45 (1982).

\bibitem{bengtssonZerwas} M. Bengtsson and P. M. Zerwas, 
{\it Phys. Lett. B\/} {\bf 208}, 306 (1988).




%%Cross Sections Ratios


%%R32 paper
\bibitem{CMS_R32_Ht} CMS Collab. (S. Chatrchyan {\it et al}.),
{\it Phys. Lett. B\/} {\bf 702}, 336 (2011).

\bibitem{CMS_R32_as} CMS Collab. (S. Chatrchyan {\it et al}.),
{\it Eur. Phys. J. C\/} {\bf 73}, 2604 (2013).



%%R32 ak7/ak5 paper 
\bibitem{CMS_ak7ak5} CMS Collab. (S. Chatrchyan {\it et al}.),
{\it Phys. Rev. D\/} {\bf 90}, 072006 (2014).



%%Jet Shapes
\bibitem{Jet_shapes_Ellis} S. D. Ellis, Z. Kunszt, and D. E. Soper,
{\it Phys. Rev. Lett.\/} {\bf 69}, 3615 (1992).

\bibitem{ATLAS_jet_shapes} ATLAS Collab. (G. Aad {\it et al}.),
{\it Phys. Rev. D\/} {\bf 83}, 052003 (2011).

\bibitem{CMS_jet_shapes} CMS Collab. (S. Chatrchyan {\it et al}.),
{\it JHEP\/} {\bf 06}, 160 (2012).

\bibitem{ATLAS_jet_shapes_ttbar} ATLAS Collab. (G. Aad {\it et al}.),
{\it Eur. Phys. J. C\/} {\bf 73}, 2676 (2013).

\bibitem{mcNLO} S. Frixione {\it et al}, 
{\it JHEP\/} {\bf 01}, 053 (2011).



%%Jet Event Shapes

\bibitem{ATLAS_event_shapes} ATLAS Collab. (G. Aad {\it et al}.),
{\it Eur. Phys. J. C\/} {\bf 72}, 2211 (2012).

\bibitem{CMS_event_shapes1} CMS Collab. (S. Chatrchyan {\it et al}.),
{\it Phys. Lett. B\/} {\bf 699}, 48 (2011).

\bibitem{CMS_event_shapes2} CMS Collab. (S. Chatrchyan {\it et al}.),
{\it JHEP\/} {\bf 10}, 087 (2014).

\bibitem{Event_shapes_phen1} A. Banfi, G. P. Salam, and G. Zanderighi, 
{\it JHEP\/} {\bf 08}, 062 (2004).

\bibitem{Event_shapes_phen2} A. Banfi, G. P. Salam, and G. Zanderighi, 
{\it JHEP\/} {\bf 06}, 038 (2012).







\end{thebibliography}
\end{document}